\documentclass[10pt,aps,prd,onecolumn,superscriptaddress,showpacs,showkeys,nofootinbib,floatfix,preprintnumbers]{revtex4}

\usepackage{appendix}
\usepackage{epsfig}
\usepackage{graphicx}
\usepackage{amsfonts}
\usepackage{amsmath}
\usepackage{psfrag}
\usepackage{amssymb,amsmath,amsthm}
\usepackage{verbatim}
\usepackage{soul}
\usepackage[normalem]{ulem}
\usepackage{color}
\usepackage{multirow}
\usepackage{titlesec}

\begin{document}

%------------------------------------------------------------------
% The following line should be uncommented if the LaTeX file is uploaded to arXiv.org
%\pdfoutput=1

%=================================================================
% Add packages and commands here. The following packages are loaded in our class file: fontenc, calc, indentfirst, fancyhdr, graphicx, lastpage, ifthen, lineno, float, amsmath, setspace, enumitem, mathpazo, booktabs, titlesec, etoolbox, amsthm, hyphenat, natbib, hyperref, footmisc, geometry, caption, url, mdframed, tabto, soul, multirow, microtype, tikz
%\usepackage{sidecap}

%=================================================================
%% Please use the following mathematics environments: Theorem, Lemma, Corollary, Proposition, Characterization, Property, Problem, Example, ExamplesandDefinitions, Hypothesis, Remark, Definition
%% For proofs, please use the proof environment (the amsthm package is loaded by the MDPI class).

%=================================================================
% Full title of the paper (Capitalized)
\title{Chirally improved quark Pauli blocking in nuclear matter and\\ applications to quark deconfinement in neutron stars}

% Authors, for the paper (add full first names)
\author{David Blaschke} 
\affiliation{Institute of Theoretical Physics, University of Wroclaw, PL-50204 Wroclaw, Poland}
\affiliation{Bogoliubov Laboratory of Theoretical Physics, Joint Institute for Nuclear Research, 141980 Dubna, Russia}
\affiliation{National Research Nuclear University (MEPhI), 115409 Moscow, Russia}
\author{Hovik Grigorian} 
\affiliation{Laboratory for Information Technologies,
	Joint Institute for Nuclear Research,
	Joliot-Curie street 6,
	141980 Dubna, Russia}
\affiliation{Department of Physics, 
	Yerevan State University, 
 	Alek Manukyan street 1, 
 	0025 Yerevan, Armenia}
\affiliation{Computational Physics and IT Division, 
	A.I. Alikhanyan National Science Laboratory, 
	Alikhanyan Brothers street 2, 
	0036 Yerevan, Armenia}
\author{Gerd R\"opke} 
\affiliation{National Research Nuclear University (MEPhI), 115409 Moscow, Russia}
\affiliation{Institute of Physics, University of Rostock, 18059 Rostock, Germany}

% Contact information of the corresponding author
%\corres{Correspondence: david.blaschke@uwr.edu.pl}

% Abstract (Do not insert blank lines, i.e. \\) 
\begin{abstract}
The relativistic mean field (RMF) model of the nuclear matter equation of state has
been modified by including the effect of Pauli-blocking owing to quark exchange 
between the baryons. 
Different schemes of a chiral enhancement of the quark Pauli blocking have been suggested
according to the adopted density dependence of the dynamical quark mass.
The resulting equations of state for the pressure are compared to the RMF model DD2 with 
excluded volume correction.
On the basis of this comparison a density-dependent nucleon volume is extracted which 
parametrises the quark Pauli blocking effect in the respective scheme of chiral enhancement. 
The dependence on the isospin asymmetry is investigated and the corresponding 
density dependent nuclear symmetry energy is obtained in fair accordance with 
phenomenological constraints.
The deconfinement phase transition is obtained by a Maxwell construction 
with a quark matter phase described within a higher order NJL model.
Solutions for rotating and nonrotating (hybrid) compact star sequences are
obtained which show  the effect of high-mass twin compact star solutions for the rotating case. 
\end{abstract}

% Keywords
\maketitle
% The fields PACS, MSC, and JEL may be left empty or commented out if not applicable
%\PACS{J0101}
%\MSC{}
%\JEL{}

%\preto{\abstractkeywords}{\nolinenumbers}

%\begin{document}
	
%%%%%%%%%%%%%%%%%%%%%%%%%%%%%%%%%%%%%%%%%%
\section{Introduction}
The behavior of baryons in a dense, strongly interacting medium and the resulting properties of dense
baryonic matter are highly interesting questions because of their relevance for explaining the interior of
compact astrophysical objects like pulsars and their mergers as well as heavy-ion collision experiments
in the NICA-FAIR energy range. 
The main problem which awaits a better theoretical formulation and
understanding is to treat the baryon as a bound state of quarks and to study the effects of this quark
substructure as a function of density. 
In particular, it is expected that at a critical value of the density the
many-baryon system will change its character and get transformed to the new state of deconfined quark
matter. 
Already before this transition occurs, the effective interaction between baryons will be strongly
modified due to the fact that the effects of quark exchange between different baryons need to be taken
into account as a requirement following from the Pauli principle on the quark level of description.
These quark substructure effects will eventually dominate over other effects due to, e.g., 
the meson exchange interaction. 
The resulting quark exchange contribution to the baryon selfenergy will entail an increase of the energy per baryon
and thus lead to a stiffening of dense baryonic matter. 
On the other hand, the quark exchange between two baryons involves already a six-quark wave function 
which is a partial delocalization of quarks and can be seen as a precursor of the transition to deconfined 
(i.e. delocalized) quark matter. 
In this transition from a many-baryon to a many-quark system the matter shall be effectively softened, 
due to the appearance of a mean field emerging from the attractive two-quark interactions. 
The value of the critical density for deconfinement is crucial for applications in heavy-ion collisions and 
compact stars.

The aspect we want to consider in this work is to investigate the influence of quark exchange on the
selfenergies of baryons and the equation of state of dense baryonic matter on the one hand and on the
phase transition to delocalized quark matter on the other. 
For quantitative estimates we shall employ a relativistic mean field theory for baryonic matter 
(the linear Walecka model) 
as well as for quark matter (the NJL-type model with higher order quark interactions) and superimpose 
the quark exchange contribution to the baryon self energy obtained within a nonrelativistic quark potential 
model of the baryon structure and the six-quark wave function.
The quark mass in this calculation has a density dependence (even inside the baryon) and is taken, e.g.,
from the NJL model calculation. 

The effect of Pauli blocking in systems of composite particles can be discussed from the quark and
nuclear level to that of atomic clusters. 
The relationship between Pauli blocking and excluded volume is known from the fact that the hard-sphere 
model of molecular interactions is based on the electron exchange interaction among atoms 
(see, e.g., Ebeling et al. \cite{Ebeling:2008mg}) which is captured, e.g., in the Carnahan-Starling EoS \cite{Carnahan:1969}.
Note that the Carnahan-Starling form of the EoS 
for multicomponent mixtures \cite{Mansoori:1971} 
has recently been reproduced for a hadron resonance gas model with induced surface tension when the packing 
fraction is not too large \cite{Sagun:2017eye}.
A recent application of the Pauli blocking effect has been found in Ref.~\cite{Ropke:2018ewt} where its role for explaining the ionization potential 
depression accessible in high-pressure experiments with warm dense plasmas has been demonstrated.
The temperature-, density- and momentum dependence of the Pauli blocking depends on the generic form of bound state wave functions and therefore concepts developed for atomic systems could thus be taken over to the case of dense hadronic systems.
Detailed parametrizations of the Pauli shift for nuclear clusters in warm, dense nuclear matter are given in \cite{Ropke:2011tr} (see also references therein). 
In Ref.~ \cite{Ropke:1986qs} it has been demonstrated that the repulsive part of effective density-dependent nucleon-nucleon interactions of the Skyrme type 
(e.g., the one by Vautherin and Brink \cite{Vautherin:1971aw}) can be reproduced by the quark exchange interaction between nucleons. 

On the other hand, for the description of repulsive interactions in dense hadronic systems the concept of an excluded volume has been 
successfully applied \cite{Bugaev:2018uum} and extended to the case of light nuclear clusters \cite{Bugaev:2020sgz}, but this application requires a medium dependence of the excluded volume parameter \cite{Vovchenko:2017drx} and thus hints to a microscopic origin from the composite nature of hadrons and clusters. 
We will therefore use the present approach to quantify such a relationship between quark Pauli blocking in dense nuclear matter and the medium
dependence of the excluded volume parameter by comparing the EoS of the present approach to the relativistic mean field approach DD2 with excluded volume \cite{Typel:2016srf}.
We would like to point out that the inclusion of the Pauli-blocking effect within a quantum statistical description of light cluster formation and dissociation in nuclear matter at subsaturation densities \cite{Ropke:1982ino,Ropke:1983lbc,Typel:2009sy}
has important consequences for the equation of state and the composition of matter as seen, e.g., in the nuclear symmetry energy
\cite{Natowitz:2010ti} that is successfully compared to experiements and in the description of supernova matter \cite{Sumiyoshi:2008qv,Ropke:2020peo} where otherwise excluded volume approaches are commonly used \cite{Lattimer:1991nc}. 

In the present investigation, we will also consider the role that the quark exchange interaction can play for
the nuclear symmetry energy. 
Here the interesting question arises inasmuch the quark exchange contribution can make the contribution from 
isovector meson exchange obsolete.
Our present study suggests that the $\rho$-meson mean field may not have any contribution for densities 
up to the onset of baryon dissociation.

Recently, the question of the softening of dense baryonic matter due to the appearance of strange
baryons became very popular and led to the hyperon puzzle: a lowering of the maximum mass of
compact stars so that the existence of pulsars with masses as high as $2~M_\odot$ could not be explained.
In principle, the approach can be extended to obtain results on the baryon self energy shifts also in the case 
that the strange quark flavor will be included. 
In that case the presented approach can make a contribution to solving the question: 
Which effect will dominate when increasing the density: the occurrence of strange baryons or of deconfined 
strange quark matter?
In this present work we want to consider as a first step only the question of non-strange quark-nucleon matter.

\section{Quark exchange in nuclear matter}

\subsection{Quark substructure effect on the selfenergy of the nucleons}

The quark substructure of nucleons becomes apparent for higher densities,
when the nucleon wave functions have a finite overlap so that the effects of
quark exchange between nucleons due to the Pauli principle on the quark level 
are no longer negligible. 
A quantitative estimate for this effect has been made within a potential model
for the nucleons as three-quark bound states \cite{Ropke:1986qs,Blaschke:1988zt},
see the Appendix for details of the derivation, and we will employ the resulting 
contribution to the nucleon selfenergy as the basis for our work.
The result has been obtained in the form of a Pauli blocking  energy shift 
 for a nucleon with momentum $P$, given spin and the isospin projection $\tau=n,p$
in nuclear matter at $T=0$,
\begin{eqnarray}
%\Sigma_{\nu}(p,p_{Fn},p_{Fp}) 
\Delta E_{\tau P}^{\rm Pauli}(P_{F,n},P_{F,p})
%\sum_{\nu'}\Omega\int_{|{\bf P}'| < P_F} \frac{d^3 {\bf P}'}{(2 \pi)^3} \Delta E_{nn'}^{\rm Pauli}
= \sum_{\tau'=n,p}\sum_{\alpha=1,2}c_{\tau \tau'}^{(\alpha)}W_\alpha (P,P_{F,\tau'})~,
%& = & \sum_{\nu'=\{n,p\}}\sum_{\alpha=1,2}C_{\nu\nu'}^{(\alpha)}W_{\alpha}(p_{F\nu'},p)~,
\label{PSE}
\end{eqnarray}
where $P_{F,\tau}$ is the Fermi momentum of a medium nucleon with the isospin projection 
$\tau=n, p$, which is directly related to the medium density by $P_{F,\tau}=(3\pi^2\, n_\tau)^{1/3}$.
The coefficients for the $c_{\tau \tau'}^{(\alpha)}$ are given in table~\ref{coef},
and their superscript index 
$\alpha=1,2$ indicates whether they apply for the one-quark or the two-quark
exchange in the two-nucleon system. 
The functions $W_{\alpha}(P,P_{F\tau'})$ are the contributions due
to the Pauli-shift in the energy spectrum of three quark bound states.
Their analytic derivation  
within a harmonic oscillator confinement model for the ground state nucleons 
according to the Ref.~\cite{Blaschke:1988zt} is detailed in the Appendix.
The resulting expression for $W_{\alpha}(P,P_{F\tau'})$ is 
\begin{eqnarray}
W_{\alpha}(P,P_{F\tau'})  =\frac{9\sqrt{3}}{64 \sqrt{\pi}}\frac{b}{m}\frac{1}{\lambda_{\alpha}^{3}} 
& &\bigg\{12\sqrt{\pi}\left[\mathrm{erf}	\left(\lambda_{\alpha}(P_{F\tau'}-P)\right)+{\rm \mathrm{erf}}\left(\lambda_{\alpha}(P_{F\tau'}+P)\right)\right]
	\nonumber\\
& & + \frac{1}{\lambda_{\alpha} P}\left\{\left[11-2\lambda_{\alpha}^{2}~P_{F\tau'}(P_{F\tau'}+p)\right]\mathrm{e}^{-\lambda_{\alpha}^{2}(P_{F\nu'}+P)^{2}}	
\right.	\nonumber\\
& & \hspace{1cm} -  \left.   \left[11-2\lambda_{\alpha}^{2}~P_{F\tau'}(P_{F\tau'}-P)\right]\mathrm{e}^{-\lambda_{\alpha}^{2}(P_{F\tau'}-P)^{2}}\right\}\bigg\}~.
%\nonumber\\
\end{eqnarray}
Here $m$ % $m=350$ MeV 
is the constituent quark mass and $b$ is the width parameter of the nucleon wave function that describes the quark substructure by a product of two 
Gaussian functions of the relative (Jacobi) coordinates in the three-quark system with $b^{-2}=\sqrt{3}m\omega$;% and the ground state energy ;
%which for $b=0.6$ fm is corresponding to the $\omega=178.425$ MeV; 
$\lambda_{\alpha}={b\alpha}/({2\sqrt{3}})$ denote the ranges for one- and two-quark exchange processes.
Values for $b$ and $\omega$ are given below.

\begin{table}[!thb]
	\begin{tabular}{|c||c|c|}
		\hline 
		 & $c_{n~\tau}^{(1)}$  & $c_{n~\tau}^{(2)}$ \tabularnewline 
		 $\tau$ &&\tabularnewline
		\hline 
		&& \tabularnewline
		n  & $\frac{15}{81}$  & $-\frac{16}{81}$ \tabularnewline &&\tabularnewline
		p  & $\frac{12}{81}$  & $-\frac{14}{81}$ \tabularnewline && \tabularnewline
		\hline 
	\end{tabular}
	\caption{Quark exchange coefficients $c_{n\tau}^{(\alpha)}$ in spin-flavor-color space.
%	corresponding to the table in Appendix (note that a factor 2 for spin degeneracy of the considered nucleon has been absorbed into these coefficients, to make them compatible with the previous (Barnes coefficients) table). 
These coefficients %$c_{\nu\nu'}^{(\alpha)}$ 
entail the symmetry relation $\Delta E^{\rm Pauli}_{pP}(P_{F,n},P_{F,p})=\Delta E^{\rm Pauli}_{nP}(P_{F,p},P_{F,n})$.}
	\label{coef} 
\end{table}

\begin{figure}[!th]
	\includegraphics[scale=0.45]{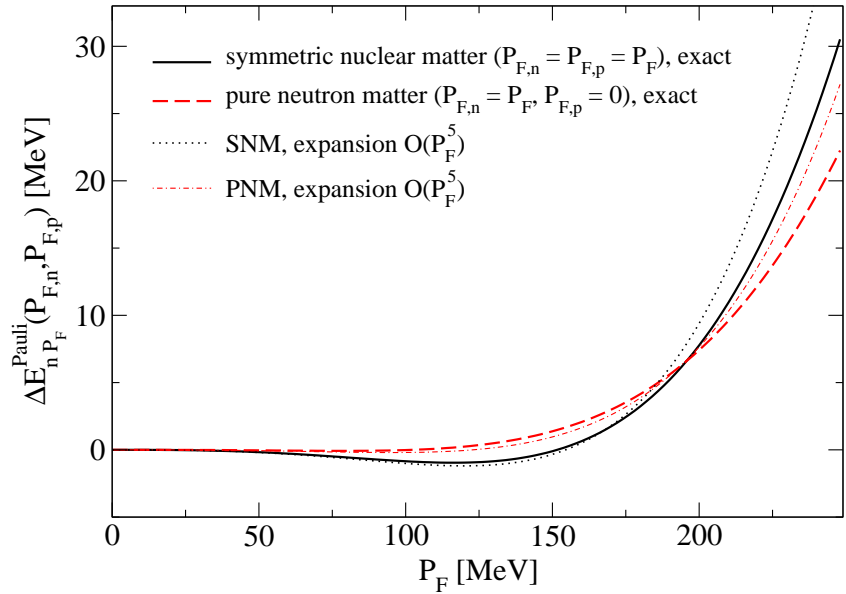}
	\caption{Exact Pauli-shifts for symmetric nuclear matter (solid line) and pure neutron matter (dashed line) as function of the Fermi momentum,
	together with their power law expansion up to $\mathcal{O}(P_F^5)$ (dotted and dash-dotted thin lines, resp.). 
	}
	\label{seasym} 
\end{figure}

%%%%%%%%%%%%%%%%%%%%%%%%%%%%%%%%%%%%%%%%%%%%%%%%%%%%%%%%%%%%%%%%%%%%%%

We want to consider as examples the two special cases:
\begin{enumerate} 
\item Symmetric nuclear matter (SNM), for which $P_{F,n}=P_{F,p}=P_{F}$ and
\begin{eqnarray}
\Delta E_{n P_F}^{\rm Pauli}(P_{F},P_{F})&=&c_{nn}^{(1)} W_1(\lambda_1 P_F) +c_{nn}^{(2)} W_2(\lambda_2 P_F) +c_{np}^{(1)} W_1(\lambda_1 P_F) +c_{np}^{(2)} W_2(\lambda_2 P_F).
\end{eqnarray}
The Pauli shift for protons is obtained using the symmetry relation $\Delta E^{\rm Pauli}_{pP}(P_{F,n},P_{F,p})=\Delta E^{\rm Pauli}_{nP}(P_{F,p},P_{F,n})$ 
that is encoded in the coefficients of table \ref{coef}.
With these coefficients and the low-density expansion (\ref{eq:Ppower}) up to fifth order in the Fermi momentum,
we obtain 
\begin{eqnarray}
\label{eq:PaulishiftSNM}
\Delta E_{n P_F}^{\rm Pauli}(P_{F},P_{F})&=&\frac{5}{8\sqrt{3\pi}} \frac{b}{m}  \left(-P_F^3+\frac{1054}{225} b^2  P_F^5\right).
\end{eqnarray}
This energy shift can be identified with a shift in the chemical
potential and thus be used to derive a contribution to the equation of state, see Ref. \cite{Ropke:1986qs}.

In order to give numerical results for the Pauli shift (\ref{eq:PaulishiftSNM}),
we adopt the values $m=350$ MeV and $b=0.59$ fm according to Ref. \cite{Blaschke:1988zt}
which reproduce quite well the single nucleon properties.
With the relation $P_F^3=(3 \pi^2/2) n$,
the Pauli blocking shift
can be given as a function of the nuclear matter density $\rho$
\begin{equation}
\label{eq:shift}
\Delta E^{\rm Pauli}(n)=a_1 n+a_2 n^{5/3},
\end{equation}
with $a_1^{\rm (SNM)} = -197.77$ MeV fm$^3$ and $a_2^{\rm (SNM)} = 1944.45$ MeV fm$^5$. 
As has been discussed already in \cite{Ropke:1986qs}, this density dependent energy shift is in good agreement with the repulsive part of the Skyrme 
Hartree-Fock shift in nuclear matter obtained by Vautherin and Brink \cite{Vautherin:1971aw}.
 
\item Pure neutron matter (PNM), for which $P_{F,p}=0$, $P_{F,n}=P_{F}$ and
\begin{eqnarray}
\Delta E_{n P_F}^{\rm Pauli}(P_{F},P_{F})&=&c_{nn}^{(1)} W_1(\lambda_1 P_F) +c_{nn}^{(2)} W_2(\lambda_2 P_F) ~.
\end{eqnarray}
Inserting the coefficients from table \ref{coef} and the low-density expansion (\ref{eq:Ppower}) up to fifth order in the Fermi momentum,
we obtain 
\begin{eqnarray}
\label{eq:PaulishiftPNM}
\Delta E_{n P_F}^{\rm Pauli}(P_{F},P_{F})&=&\frac{5}{24\sqrt{3\pi}} \frac{b}{m}  \left(-P_F^3+\frac{1666}{225} b^2  P_F^5\right).
\end{eqnarray}
Inserting the relation $P_F^3=3 \pi^2 n$ between Fermi momentum and density for PNM, we obtain the energy shift in the form (\ref{eq:shift}) with the 
coefficients $a_1^{\rm (PNM)} = -131.85$ MeV fm$^3$ and $a_2^{\rm (PNM)} = 3252.57$ MeV fm$^5$. 
\end{enumerate} 
Fig. \ref{seasym} shows the Pauli blocking shift  for the SNM and PNM cases as function of the Fermi momentum.

\begin{figure}[!th]
%\vspace{-1cm}
	\includegraphics[scale=0.4]{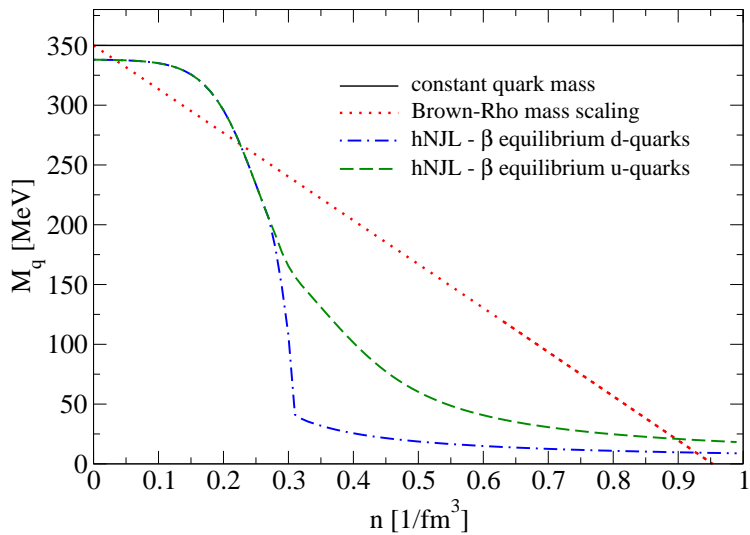}
	\caption{Dependence of the quark masses on density: constant quark mass (solid line), Brown-Rho scaling (dotted
		line), hNJL model in $\beta$-equilibrium for u-quarks (dashed line) and for d-quarks (dash-dotted line).}
	\label{qm}
\end{figure}

\subsection{Chiral improvement of the quark Pauli blocking}

One of the main shortcomings when applying the results for the quark Pauli blocking obtained within the nonrelativistic quark model to the 
equation of state of dense nuclear and neutron star matter up to the deconfinement phase transition is the fact that the quark mass is a
medium-independent constant in this model.
From chiral perturbation theory it is known that the chiral condensate $\langle \bar{q} q\rangle$ melts in a dense hadronic matter environment
so that the constituent quark mass shall be reduced towards its value in the QCD Lagrangian which obeys approximate chiral symmetry.

\begin{figure}[!hb]
	\includegraphics[scale=0.4]{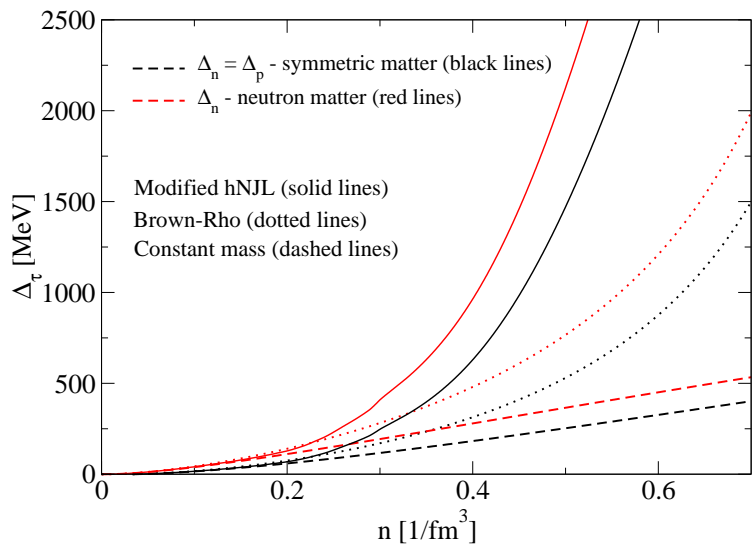}

	\caption{The Quark exchange contributions to self energy for symmetric nuclear matter (black lines)
		and for pure neutron matter (red lines) as a function of the baryon density.
		Constant quark mass case (dashed lines), Brown-Rho scaling (dotted lines)
		and hNJL model case (solid lines).}
	\label{paulishifts}
\end{figure}

Effective chiral quark models for the low-energy sector of QCD are capable of addressing the aspect of dynamical chiral symmetry breaking and 
its restoration in a hot and dense medium, but have a problem with modeling confinement of quarks in hadrons. 
Here we suggest a compromise. We will adopt a density dependence for the dynamically generated quark mass and thus achieve a chiral 
improvement of the quark Pauli blocking shift. We will discuss in the following three schemes for this density-dependent quark mass:
(i) a constant quark mass, (ii) a linear density dependence (called Brown-Rho scaling \cite{Brown:1991kk}) and 
(iii) a density dependence according to the calculation within a higher order Nambu--Jona-Lasinio model \citep{Benic:2014iaa}.
These density dependences of the quark mass are illustrated in Fig.~\ref{qm}.
In Fig.~\ref{paulishifts} we show the energy shifts $\Delta_\tau(n)=\Delta E_{\tau P_F}^{\rm Pauli}(P_{F},P_{F})$ resulting from the insertion 
of the density dependences for the Fermi momenta and the quark mass as shown in Fig.~\ref{qm} into Eq.~(\ref{eq:PaulishiftSNM}) and Eq.~(\ref{eq:PaulishiftPNM}) for SNM and PNM, respectively. 

In order to arrive at a model for dense (asymmetric) nuclear matter with quark substructure effects at supranuclear densities, we will adopt a combined 
approach consisting of a relativistic mean-field (RMF) approach to nuclear matter which in its simplest form is the well-known linear Walecka (LW) model
\cite{Chin:1974sa,Walecka:1974qa}, to which we add the repulsive quark Pauli blocking interaction which should then partly replace the vector meson exchange at high densities and play the role of a precursor of the delocalization of the nucleon wave functions in the quark deconfinement transition.
Such a combined approach has been very successfully employed before in the description of light nuclear clusters in nuclear matter by Typel et al. 
\cite{Typel:2009sy}, where selfenergy effects for nucleons were treated within a relativistic mean-field theory while the cluster formation is described within  a 
nonrelativistic quantum statistical approach that allowed to account for the reduction of the binding energy of the clusters due to nucleonic Pauli blocking, leading to the Mott dissociation  of the clusters and the formation of uniform nuclear matter around the saturation density.

In  the present work, the role of the clusters is played by the nucleons as three-quark bound states, subject to a quark Pauli blocking effect that triggers their Mott dissociation into deconfined quark matter described in a relativistic mean-field model for which we adopt the higher order Nambu--Jona-Lasinio (hNJL) model of 
Ref.~\cite{Benic:2014iaa}. At lower densities, in order to make contact with the phenomenology of nuclear matter saturation properties,  the Fermi gas model of nucleons three-quark bound states with a hard core repulsion from quark Pauli blocking has to be augmented with additional attraction and repulsion as described, e.g., by the coupling to scalar and vector mean fields in the $\sigma-\omega$ (LW) model.

\section{Equation of state of cold, dense matter with deconfinement transition}

\subsection{Relativistic mean field model with quark exchange contribution}

The modification of the LW model to account for quark exchange (Pauli blocking) effects among nucleons is introduced by additional 
contributions to the pressure ($p_{\rm ex}$) and to the energy density  ($\varepsilon_{\rm ex}$) as
\begin{eqnarray}
\label{EoS}
P & = & \frac{1}{8\pi^{2}}\sum_{\tau=n,p}\left[-E_{\tau}^{*}m_{\tau}^{*2}P_{F,\tau}+\frac{2}{3}E_{\tau}^{*}P_{F,\tau}^{3}+m_{\tau}^{*4}\log\left(\frac{\text{\ensuremath{E_{\tau}^{*}+P_{F,\tau}}}}{m_{\tau}^{*}}\right)\right]
%\nonumber\\& &
 + \frac{1}{2}G_\omega n^{2}-\frac{1}{2}G_\sigma n_{s}^{2}+P_{\rm ex}, \\
\varepsilon & = & \frac{1}{8\pi^{2}}\sum_{\tau=n,p}\left[2\, E_{\tau}^{*3}P_{F,\tau}-E_{\tau}^{*}m_{\tau}^{*2}P_{F,\tau}-m_{\tau}^{*4}\log\left(\frac{\text{\ensuremath{E_{\tau}^{*}+P_{F,\tau}}}}{m_{\tau}^{*}}\right)\right]
%\nonumber \\& & 
+ \frac{1}{2}G_\omega n^{2}+\frac{1}{2}G_\sigma n_{s}^{2}+\varepsilon_{\rm ex},
%\nonumber 
\label{EoS-e}
\end{eqnarray}
where $n=n_p+n_n$ is the baryon density, $n_{s}=n_{s,p}+n_{s,n}$ the scalar density, and for each baryon we have 
\begin{eqnarray}
n_{s,\tau} & = & \frac{m_{\tau}^{*}}{2\pi^{2}}\left[E_{\tau}^{*}P_{F,\tau}-m_{\tau}^{*2}\log\left(\frac{\text{\ensuremath{E_{\tau}^{*}+P_{F,\tau}}}}{m_{\tau}^{*}}\right)\right],\\
E_{\tau}^{*} & = & \sqrt{m_{\tau}^{*2}+P_{F,\tau}^{2}}\\
n_{\tau} & = & \frac{P_{F,\tau}^{3}}{3\pi^{2}},\\
m_{\tau}^{*} & = & m_{\tau}-G_\sigma n_{s,\tau},\\
\mu_{\tau} & = & E_{\tau}^{*}+G_\omega n_{\tau}+\mu_{{\rm ex},\tau}.
\end{eqnarray}
The effective coupling constants $G_\sigma = ({g_{\sigma}}/{m_{\sigma}})^{2}$ and $G_\omega = ({g_{\omega}}/{m_{\omega}})^{2}$
are adjusted in order to fit the saturation point of symmetric nuclear matter with the phenomenological binding energy per nucleon, see table \ref{Mcoupling}
and the section with the results. 

%$E_0=\varepsilon(n_0)/n_0 - m=-16$ MeV at the nuclear saturation density $n_0=0.15$ fm$^{-3}$.

In the relativistic mean-field EoS of Eqs.~(\ref{EoS}) and (\ref{EoS-e}) we have introduced also the contribution to the thermodynamical quantities
that originate from the quark exchange self energy via

\begin{eqnarray}
\mu_{{\rm ex},\tau} & = & \Delta_{\tau}(n,x)=\Delta E_{\tau P_{F,\tau}}^{\rm Pauli}(P_{F,n},P_{F,p}),\\
%\Sigma_{\nu}(p_{F,\nu};p_{Fn,},p_{Fp}),\\
\varepsilon_{\rm ex} & = & \int_{0}^{n}dn'\{x\Delta_{p}(n',x)+(1-x)\Delta_{n}(n',x)\},\\
P_{\rm ex} & = & \sum_{\tau=n,p}\mu_{{\rm ex},\tau}\: n_{\tau}-\varepsilon_{\rm ex},
\end{eqnarray}
where $n_p$ ($n_n$) denotes the proton (neutron) density and $x=n_p/n$ is the proton fraction. 

\subsection{NJL model with higher order quark interactions}
\label{ssec:hNJL}

In order to describe cold quark matter 
that is significantly stiffer than
the ideal gas, we employ a recently developed 
generalization of the NJL
model which includes higher order quark interactions 
in both, Dirac scalar and Dirac
vector channels (hNJL), see \citep{Benic:2014iaa} and references therein.
The thermodynamic potential density of the 2-flavor hNJL model for a homogeneous quark matter 
system in the mean-field approximation is given by
\begin{eqnarray}
\Omega &=&-2N_c \sum_{q=u,d}\left\{\int_0^{\Lambda} \frac{dp \, p^2}{2\pi^2} E_q
 - \frac{1}{16\pi^2}\left[ 
% (2\bar{\mu}_q^3 - 5 M_q^2\bar{\mu}_q)\sqrt{\bar{\mu}_q^2 - M_q^2} + 3M_q^4 \ln \left( \frac{\sqrt{\bar{\mu}_q^2 - M_q^2} + \bar{\mu}_q}{M_q}
 (\frac{2}{3}E_{F,q}p_{F,q}^{3} - M_q^2E_{F,q}p_{F,q} + M_q^4 \ln \left( \frac{E_{F,q}+p_{F,q}}{M_q}
 \right) \right] \right\}\nonumber\\
&&+ U - \Omega_0~,
\label{eq:pot}
\end{eqnarray}
where 
\begin{eqnarray}
U = \frac{g_{20}}{\Lambda^2}\sigma^2 + 
3 \frac{g_{40}}{\Lambda^8}\sigma^4-
3\frac{g_{22}}{\Lambda^8}\sigma^2\omega^2
-\frac{g_{02}}{\Lambda^2} \omega^2
-3\frac{g_{04}}{\Lambda^8}\omega^4 ~
\end{eqnarray}
is the potential energy density and the quark quasiparticle dispersion relation is
$E_q = \sqrt{{p}^2+M_q^2}$, with
\begin{eqnarray}
M_q&=&m_q+2\frac{g_{20}}{\Lambda^2}\sigma+4\frac{g_{40}}{\Lambda^8}\sigma^3
-2\frac{g_{22}}{\Lambda^8}\sigma\omega^2~,\\
%\tilde{\mu}_q 
E_{F,q}&=& \mu_q - 2\frac{g_{02}}{\Lambda^2}\omega-
4\frac{g_{04}}{\Lambda^8}\omega^3-2\frac{g_{22}}{\Lambda^8}\sigma^2
\omega~.
\end{eqnarray}
The model parameters are the 4-quark scalar and vector
couplings $g_{20}$, and $g_{02}$, the 8-quark scalar and 
vector couplings $g_{40}$ and $g_{04}$ as well as the current quark mass
$m$ and the momentum cutoff $\Lambda$ placed 
on the divergent zero-point energy.
Furthermore, the subtraction of 
the constant $\Omega_0$ ensures zero pressure in the vacuum.

The model is solved by minimizing the thermodynamic potential density
with respect to the mean-fields 
$X=\sigma, \omega$, i. e. %EoS14+hnjl
\begin{eqnarray}
\frac{\partial \Omega}{\partial X} = 0~,
\end{eqnarray}
and the pressure is obtained from the relation $P=-\Omega$.

In this work we use the parameter set of 
Ref.~\citep{Kashiwa:2006rc} with
$g_{20} = 2.104$, $g_{40} = 3.069$, $m_q=5.5$ MeV, and $\Lambda=631.5$ MeV.
The vector channel strengths are
quantified by
\begin{eqnarray}
\eta_2 = \frac{g_{02}}{g_{20}} \, , \qquad 
\eta_4 = \frac{g_{04}}{g_{40}}~.
\end{eqnarray}
We will concentrate on the parameter space 
where $\eta_2$ is small and
use $\eta_4$ to control the stiffness of the EoS.
With small $\eta_2$ we do not delay the onset of quark matter.
Additionally, we put $g_{22}=0$ \citep{Benic:2014iaa}.

This approach allows us to calculate 
partial pressures $P_q$ 
and partial densities $n_q = \partial P_q/\partial\mu_q$ for $q=u,d$.
In a cold stellar environment, the processes $d\rightarrow u + e^- + \bar{\nu}_e$
and $u + e^- \rightarrow d + \nu_e$ result in the
$\beta$- equilibrium relation for the chemical potentials $\mu_d = \mu_u + \mu_e$,
since the neutrinos leave the star and do not take part in the chemical equilibration.
Local charge neutrality requires
\begin{eqnarray}
\frac{2}{3}n_u - \frac{1}{3} n_d -n_e = 0~.
\end{eqnarray}
The total pressure in the quark phase 
is given as $P = P_u + P_d + P_e$, where
$P_e$ is the electron pressure 
given by the relativistic ideal gas formula.
The baryon chemical potential in the quark phase 
can be calculated from
\begin{eqnarray}
\mu = \mu_u + 2\mu_d~.
\end{eqnarray}
and the respective baryon number density is
\begin{eqnarray}
n = \frac{\partial P}{\partial \mu}  = \frac{n_u + n_d}{3}~.
\end{eqnarray}

\subsection{Quark deconfinement phase transition}

To construct a thermodynamically consistent 
hybrid EoS we use the Maxwell construction, which 
is tantamount to assuming a large surface tension at the 
hadron-quark interface.
The critical baryon chemical potential is obtained by
matching the pressures from the low 
density and the high density phase.
The first order phase transition obtained 
by the Maxwell construction generates a jump in the density
and the energy density, see 
also Fig.~\ref{Maxwell} below and 
the discussion in the following Section.

\begin{figure}[!thb]
	\includegraphics[scale=0.4]{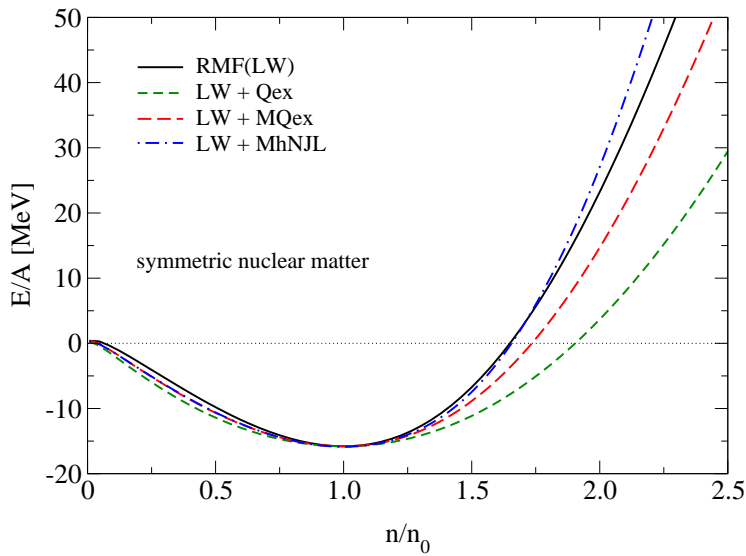}
	\caption{Energy per nucleon of symmetric matter for RMF(LW) (solid lines),
		LW+Qex (dashed lines) and LW+MQex (long-dashed lines) models.}	
	\label{FreeE}
\end{figure}

\section{Results}

\subsection{Parametrization of the model}

For the calculations we fixed the parameters of our models on the
properties of symmetric nuclear matter at the saturation density $n_{0}$ = 0.153 fm$^{-3}$, at
which the binding energy is $E/A$ = $\varepsilon/n-m_{n}$= $-15.8$ MeV.
The parameters are given in the Table \ref{Mcoupling}.

\begin{table}
	\begin{tabular}{|c|c|c|c|c|c|}
		\hline 
& $(g_{\omega}/m_{\omega})^{2}${[}fm$^{2}${]}  & $(g_{\sigma}/m_{\sigma})^{2}$ {[}fm$^{2}${]} & $K$ [MeV] & $E_s$ [MeV] & $R_{1.4}$ [km]
		\tabularnewline
		\hline 
		\hline 
		RMF (LW) & 11.6582  & 15.2883 &608.874&21.58& 13.22 \tabularnewline
		\hline 
		LW+Qex  & 6.11035  & 9.91197 &331.958&32.04&13.70\tabularnewline
		\hline 
		LW+MQex  & 8.59170  & 13.29118 &481.713&34.12&14.40\tabularnewline
		\hline 
		LW+MhNJL  & 9.25683 & 13.9474&582.831&31.55&14.29\tabularnewline
		\hline 
	\end{tabular}
	\caption{Parameter sets for vector ($G_\omega=(g_\omega/m_\omega)^2$) and scalar ($G_\sigma=(g_\sigma/m_\sigma)^2$)  meson couplings, the compressibility  $K$ and symmetry energy $E_s$ at the nuclear saturation density as well as radius $R_{1.4}$ of the neutron star with mass $1.4~M_\odot$ 
for the RMF (LW) model and for modified LW models with quark exchange contributions for different density-dependences of the quark mass: 
constant quark mass (LW+Qex), Brown-Rho scaling (LW+MQex) and hNJL model (LW+MhNJL).}
	\label{Mcoupling} 
\end{table}

In Fig. \ref{FreeE} we demonstrate the properties of symmetric nuclear
matter as a function of the baryon density. As it can be seen from
the values of coupling constants of mesons in Table \ref{Mcoupling}
in comparison with those of the LW model the repulsion of the $\omega$-meson is
partially replaced by the inclusion of Pauli- blocking via quark exchange
mechanism. 
At low densities the binding energy per baryon goes to zero since no nuclear cluster
formation is included here. For a detailed discussion of this aspect, see \cite{Typel:2009sy,Natowitz:2010ti}.

\subsection{Equation of State }

\begin{figure}[!thb]
	\includegraphics[scale=0.3]{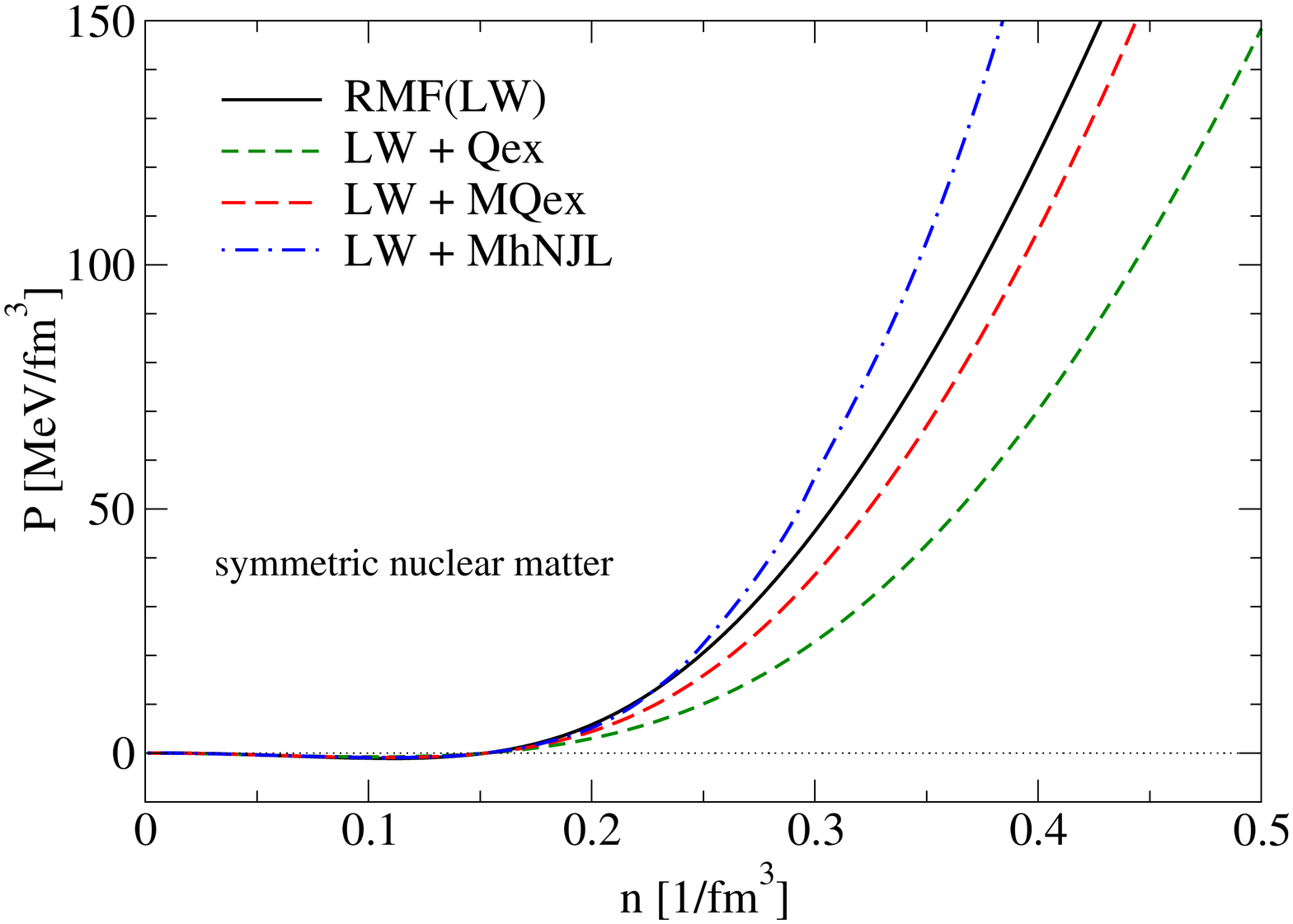}
	\includegraphics[scale=0.3]{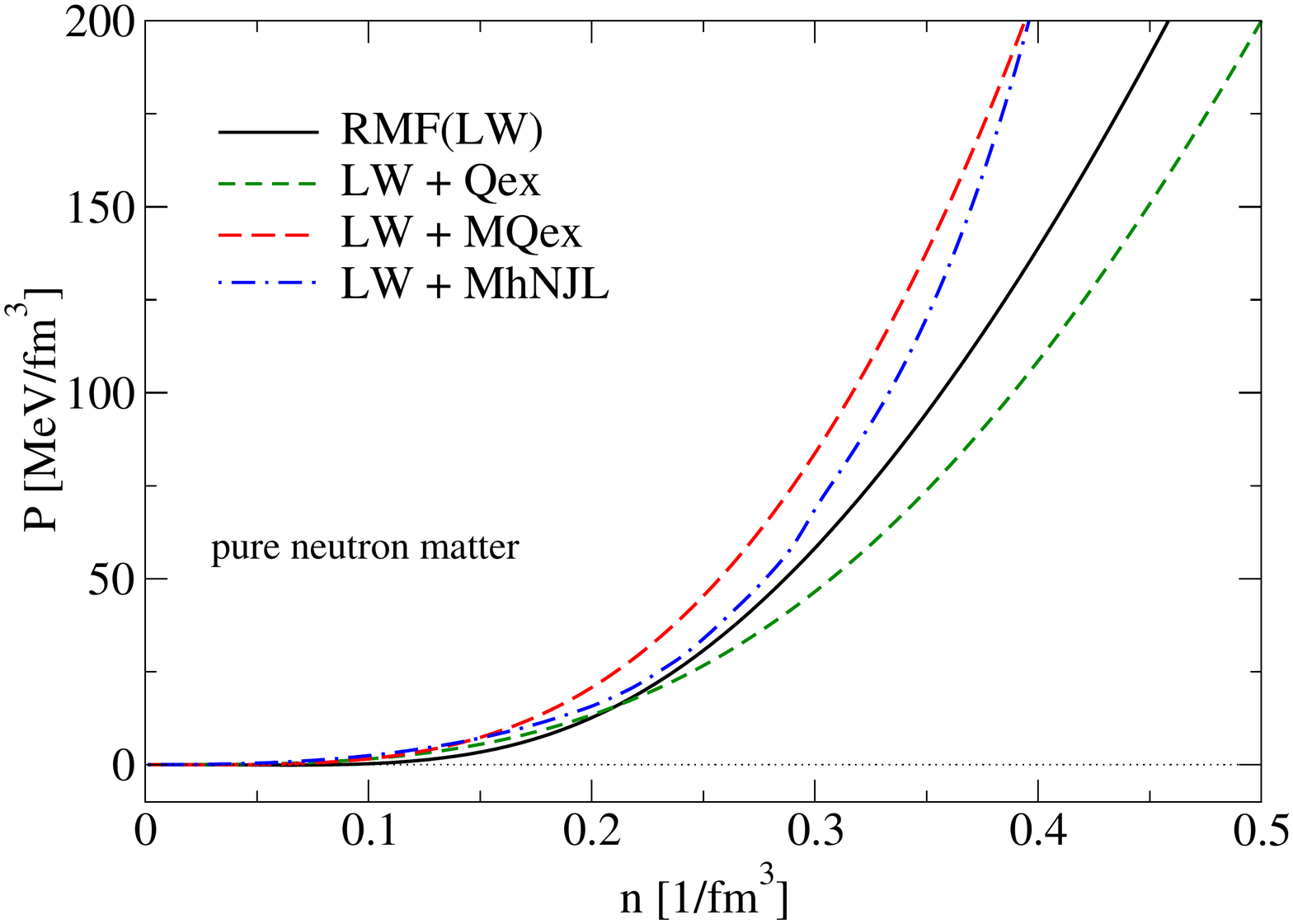}	
	\caption{Left panel: Pressure as a function of number density for symmetric nuclear matter.
	Right panel: EoS for pure neutron matter for all models.
		\label{EoSLWP} }
\end{figure}

The EoS for the nuclear matter is obtained and in Fig.~\ref{EoSLWP} the pressure as a function of the density 
is shown for symmetric matter (left panel) and for pure neutron matter (right panel).
The symmetry energy is shown below in Fig.~\ref{SymEn} for all our models. 

\subsection{Comparison with nucleonic excluded volume}

\begin{figure}[!thb]
	\includegraphics[scale=0.4]{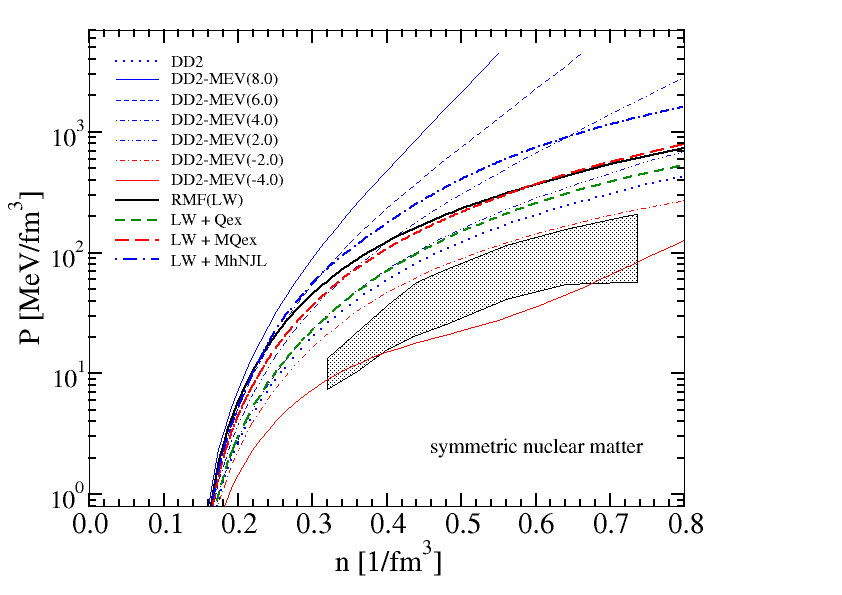}
	\caption{Same as the left panel of Fig.~\ref{EoSLWP} on a logarithmic pressure scale and 
	compared to a set of EoS for the DD2 EoS with excluded volume corrections from Ref.~\cite{Typel:2016srf}.
	From this comparison one could read-off a density dependent excluded volume corresponding 
	to the quark Pauli blocking effect in the equation of state. The hatched region corresponds to the 
	constraint derived from the analysis \cite{Danielewicz:2002pu}  of flow data from heavy ion collision experiments.}
	\label{EoS-Vex} 
\end{figure}

\begin{figure}[!thb]
	\includegraphics[scale=0.4]{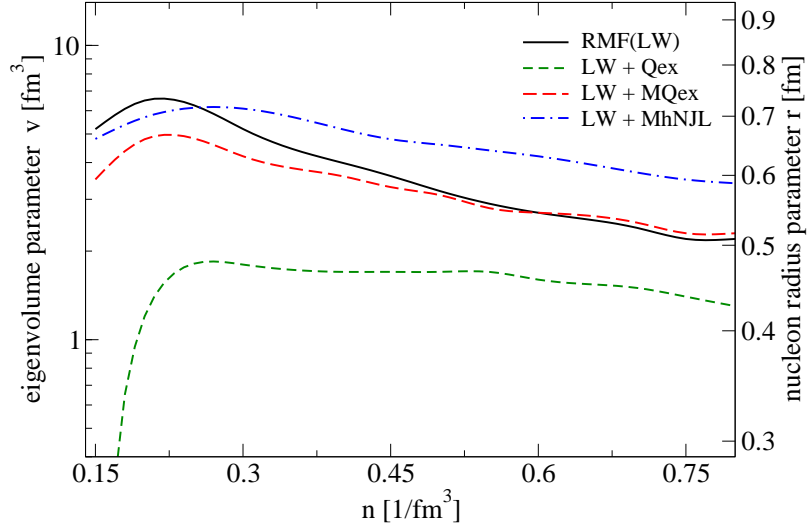}
	\caption{Density dependence of the nucleonic eigenvolume parameter ${\rm v}$ that would reproduce the 
	quark Pauli blocking EoS of the present approach for the DD2 EoS with excluded volume from Ref.~\cite{Typel:2016srf}. 
	Identifying the eigenvolume parameter with the van-der-Waals excluded volume ${\rm v}=16 \pi r^3/3$ one can extract
	a nucleon radius parameter $r$ shown on the alternative axis. 
	}
	\label{fig:Exvol} 
\end{figure}

It is interesting to compare the effect of accounting for the compositeness and finite extension of the nucleon wave function 
by the Pauli blocking effect with the phenomenological improvement of nuclear matter models by implementing a nucleonic excluded 
volume like in the van-der-Waals gas.
In Fig.~\ref{EoS-Vex} we show the equations of state for pressure vs. density that results from our LW model with 
chirally enhanced Pauli blocking and the RMF model DD2 for different values of the nucleonic excluded volume parameter \cite{Typel:2016srf}.
Also shown is the flow constraint derived from the analysis of heavy ion collision experiments \cite{Danielewicz:2002pu}.
From comparing the Pauli-blocking improved LW models with excluded-volume corrected DD2 models, 
we have extracted a density-dependent excluded volume parameter and the corresponding hard-core radius for nucleons.
We note that these results compare very well with nucleonic hard-core radii obtained within the induced surface tension approach 
reported in \cite{Bugaev:2018uum}. 
A thorough analysis of the critical temperature of  symmetric nuclear matter, the incompressibility of the  normal nuclear  
matter and the proton flow constraint  clearly shows  \cite{Bugaev:2018uum,Ivanytskyi:2017pkt} that a hard-core radius of nucleons 
up to 0.45 fm is still consistent with the available experimental data. 
Therefore, the short dashed curve in Fig.~\ref{fig:Exvol}  is perfectly consistent with the known symmetric nuclear matter properties.

It should be stressed that smaller values of $r\approx 0.35$ fm for the nucleon hard-core radius
are obtained from fits to heavy-ion collision data for hadron production at LHC and RHIC energies so that a dependence of the 
hard-core radius on the chemical freezeout temperature was conjectured in  \cite{Bugaev:2018uum}.
Extending the present approach to finite temperatures, such a temperature dependence is expected to result from the 
temperature dependence of the quark Pauli-blocking energy shift.

\subsection{Applications for neutron stars}

\begin{figure}[!thb]
\vspace{-5mm}
	\includegraphics[scale=0.3]{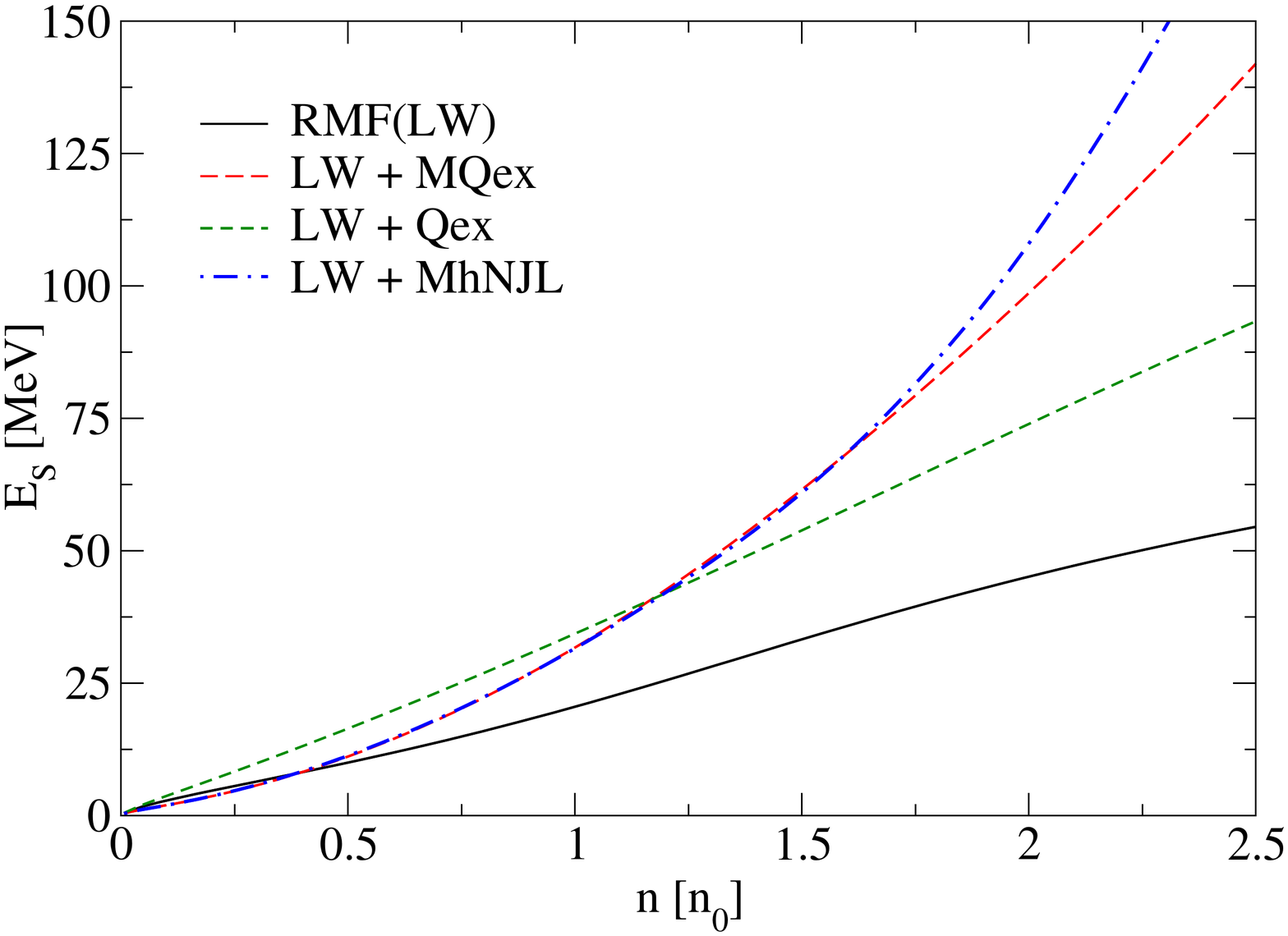}
	\includegraphics[scale=0.3]{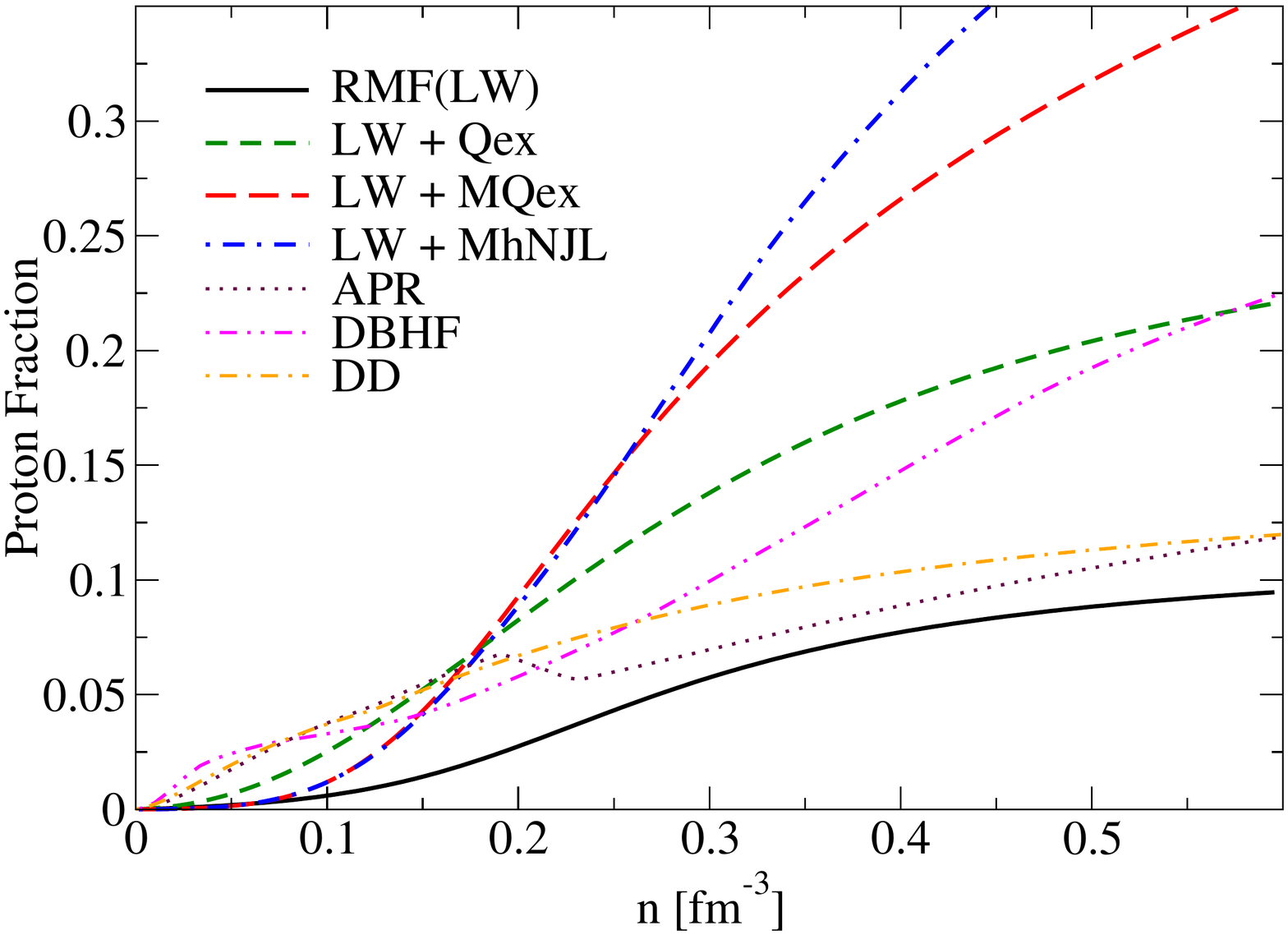}
	\caption{Left panel: Symmetry energy as a function of density for RMF(LW) (solid lines),
		LW+Qex (dashed lines), LW+MQex (long-dashed lines) and LW+MhNJL models.
		Right panel: Proton fraction as a function of density for these models (same line styles) 
		in comparison to standard neutron star EoS: APR (dotted line), DBHF (dash-double dotted line) and the DD RMF model (short dash-dotted line).}	
	\label{SymEn}
\end{figure}

In the left panel of Fig.~\ref{SymEn} we show the symmetry energy as a function of the density for all considered models. 
In the right panel of Fig.~\ref{SymEn} we show the proton fraction as a function of the density which results from 
accounting for the $\beta-$ equilibrium with electrons for all considered models. 

\begin{figure}[!h]
\vspace{-10mm}
	\includegraphics[scale=0.3]{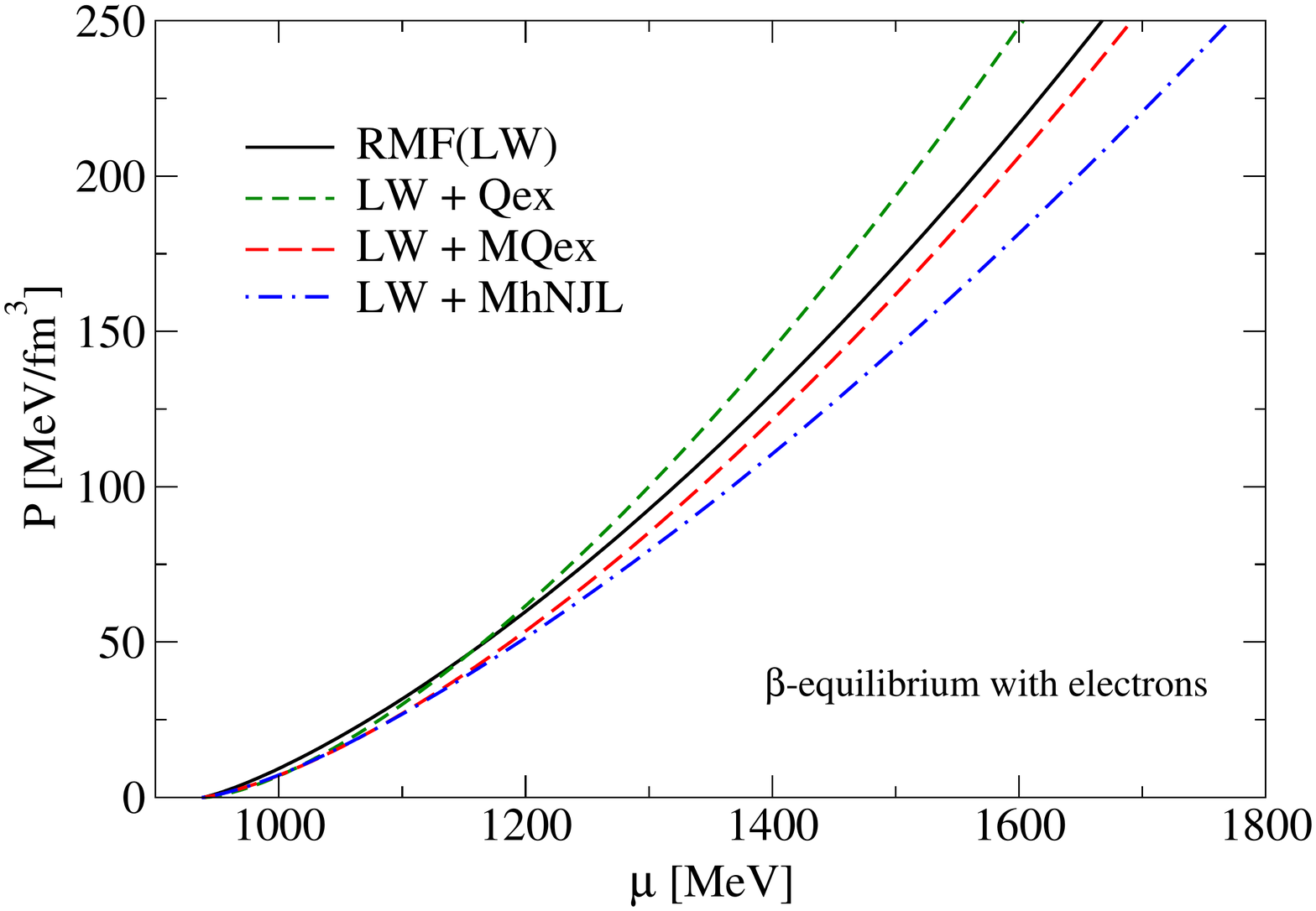}
	\includegraphics[scale=0.3]{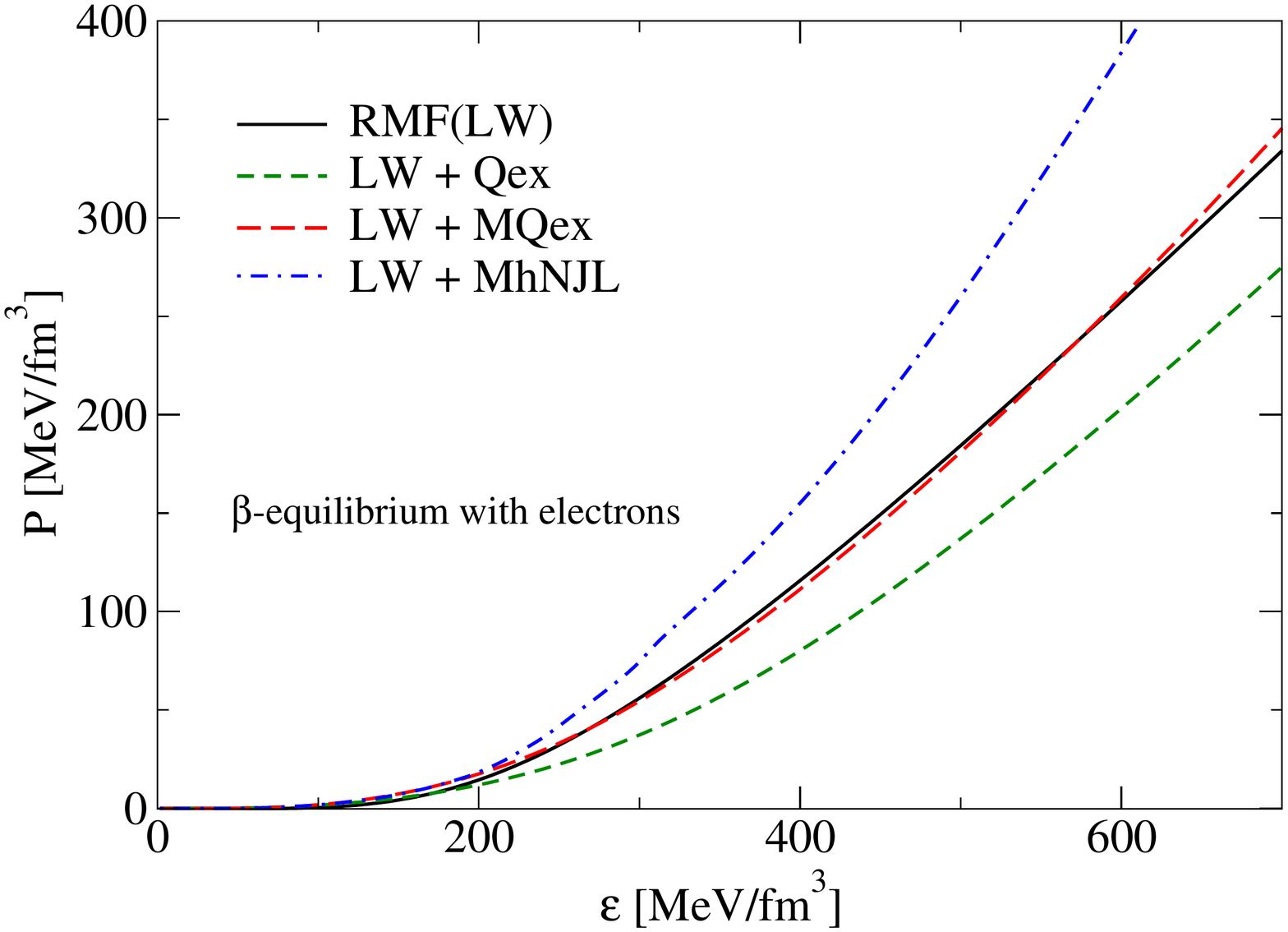} \\{}
	\vspace{-10mm}
	
	\includegraphics[scale=0.3]{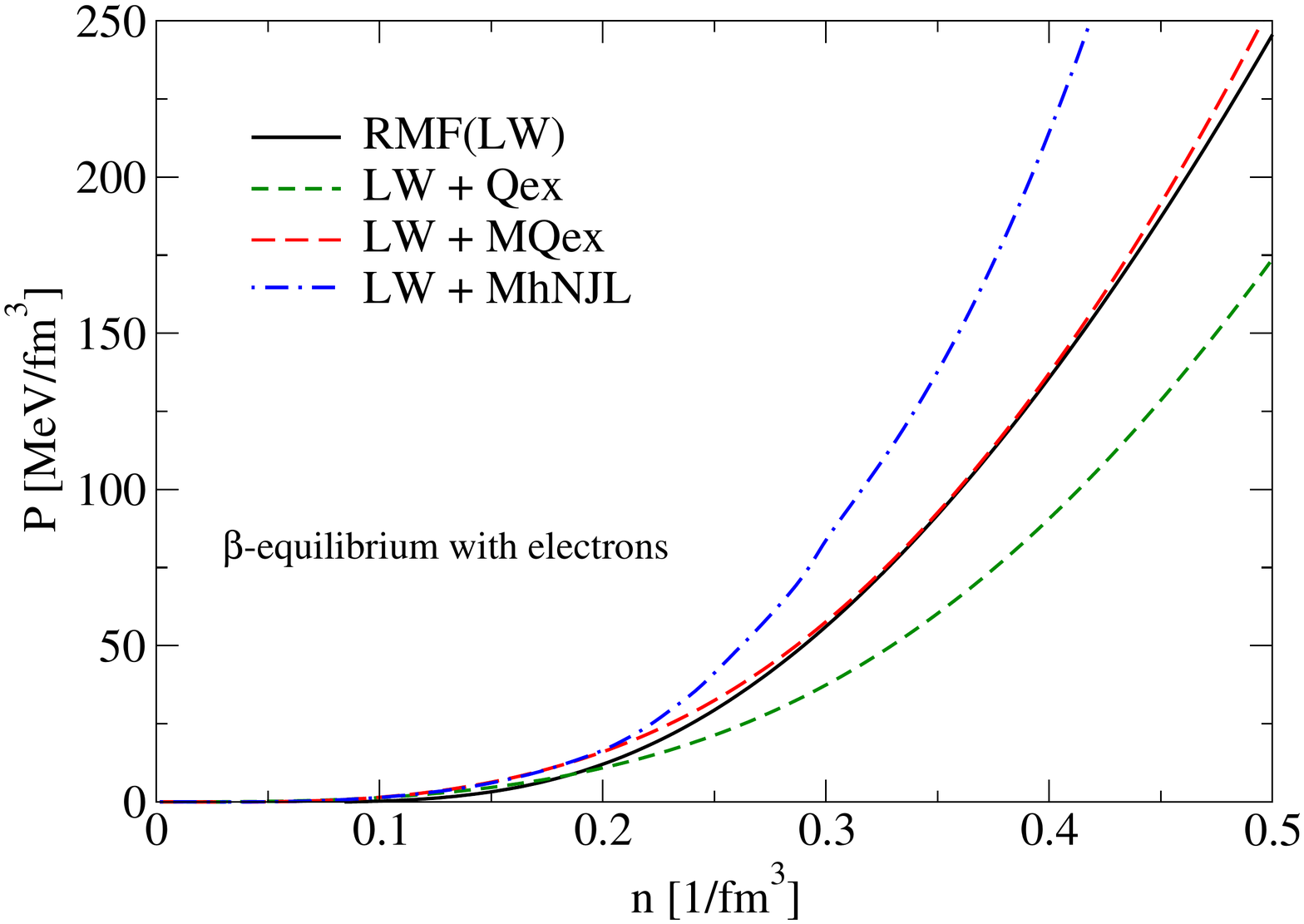}
	\includegraphics[scale=0.3]{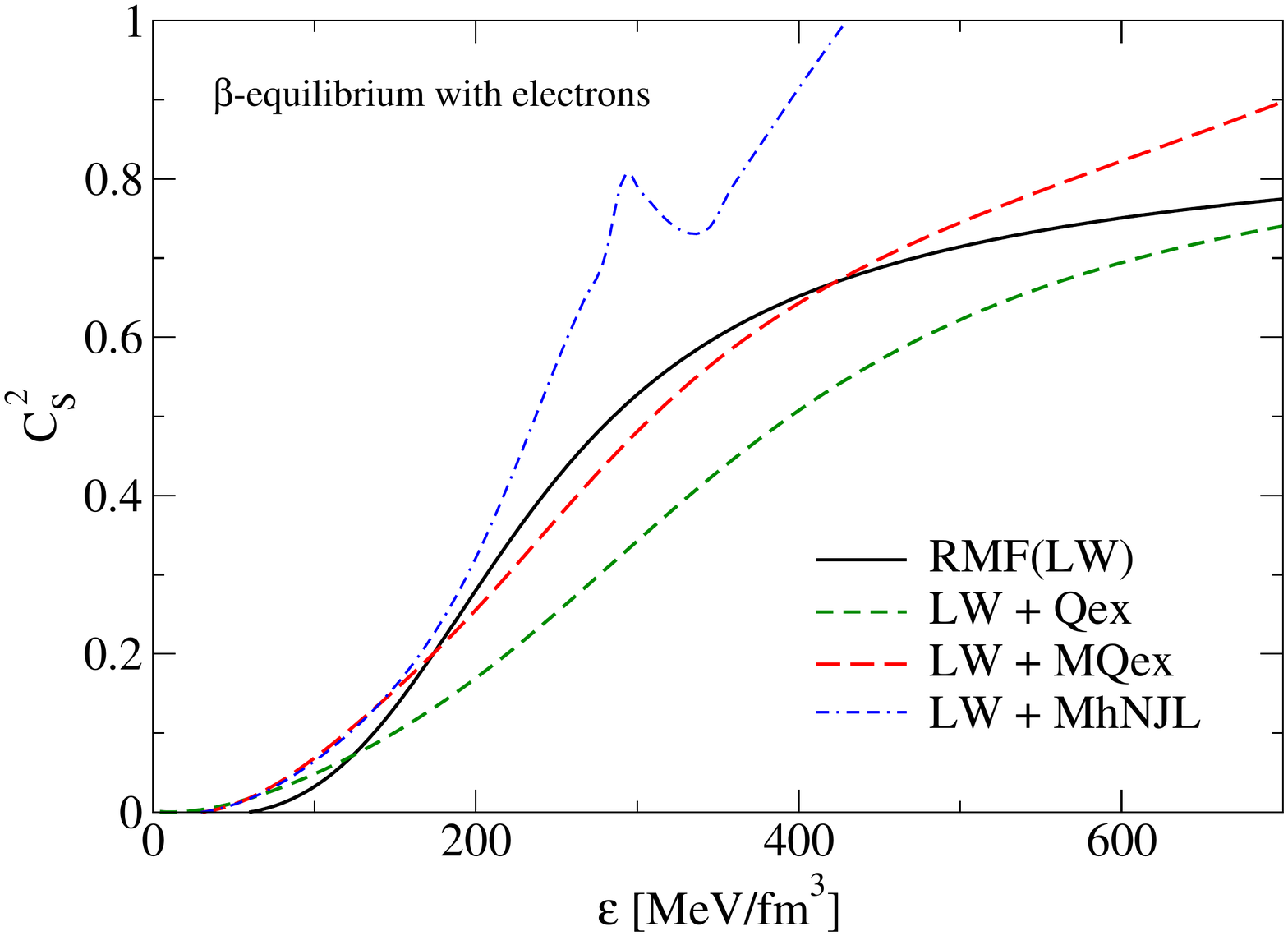}	
	\caption{EoS in $\beta-$equilibrium with electrons for all considered models. 
	Upper left panel: pressure as function of baryo-chemical potential $\mu$; 
	lower left panel: pressure as function of baryon density $n=dP/d\mu$;
	upper right panel: pressure as function of energy density $\varepsilon=-P+\mu n$;
	lower right panel: squared speed of sound $c_s^2=dP/d\varepsilon$.}	
	\label{BeqEoS}
\end{figure}

For all three models we construct the thermodynamics of stellar matter in $\beta-$equilibrium fulfilling the charge neutrality condition
with electrons and protons. 
In Fig.~\ref{BeqEoS} we show the EoS in $\beta-$equilibrium for all considered models in comparison with the LW EoS.

We consider the problem of causality in our modeling and show in the lower right panel of Fig.~\ref{BeqEoS} 
the dependence of the squared speed of sound on density. 
As it is shown for models LW, LW+Qex and LW+MQex for all relevant densities the causality holds since $c_s^2<1$. 
For the model LW+MhNJL the causality is violated
for high densities where already the transition to quark matter has to happen. 
This fact is consistent with our modeling because as a
mass function for the quarks we took the behavior corresponding to
hNJL model, see Fig.~\ref{qm}.

\begin{figure}[!th]
\vspace{-5mm}
	\includegraphics[scale=0.3]{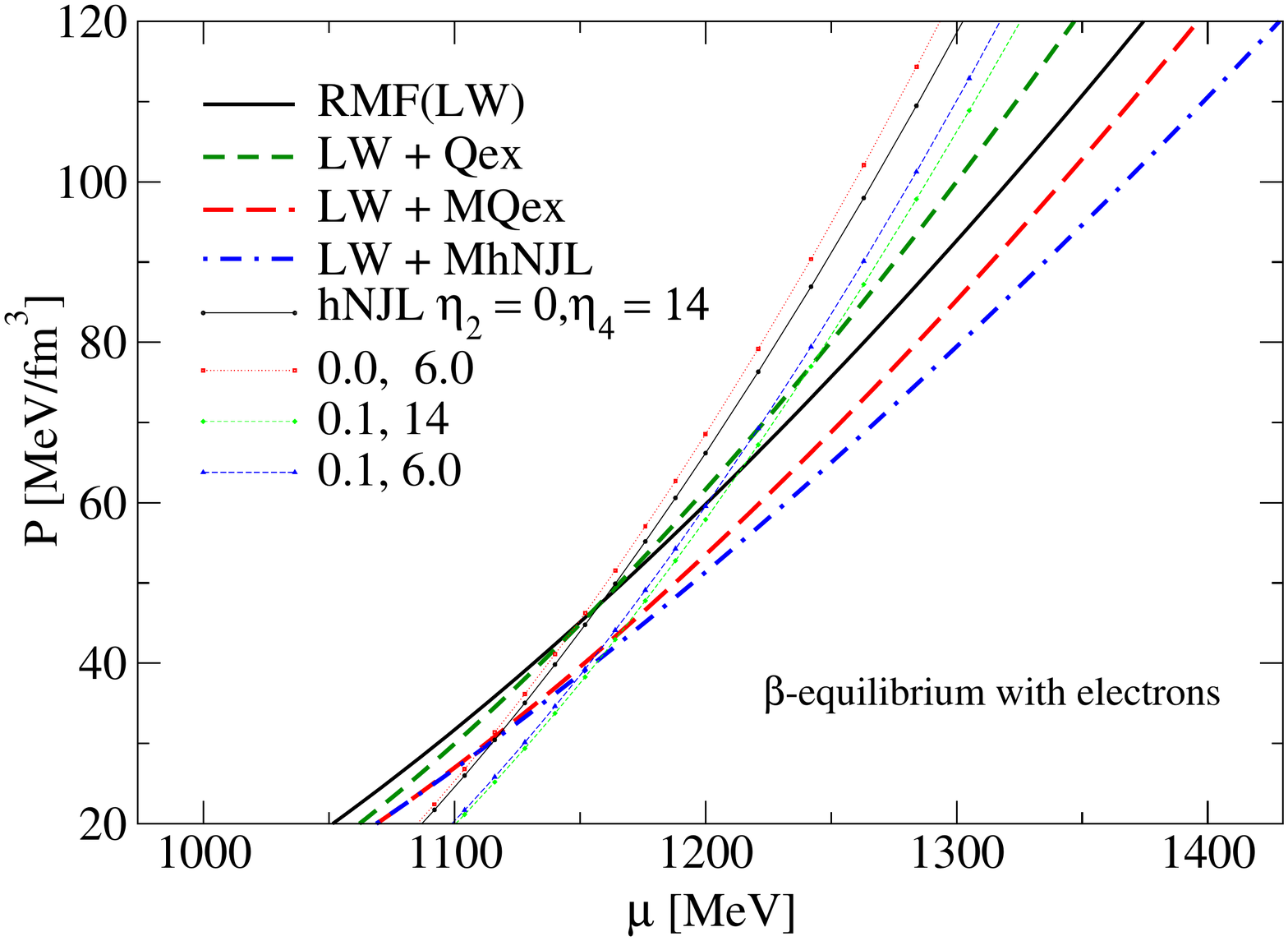}
	\includegraphics[scale=0.3]{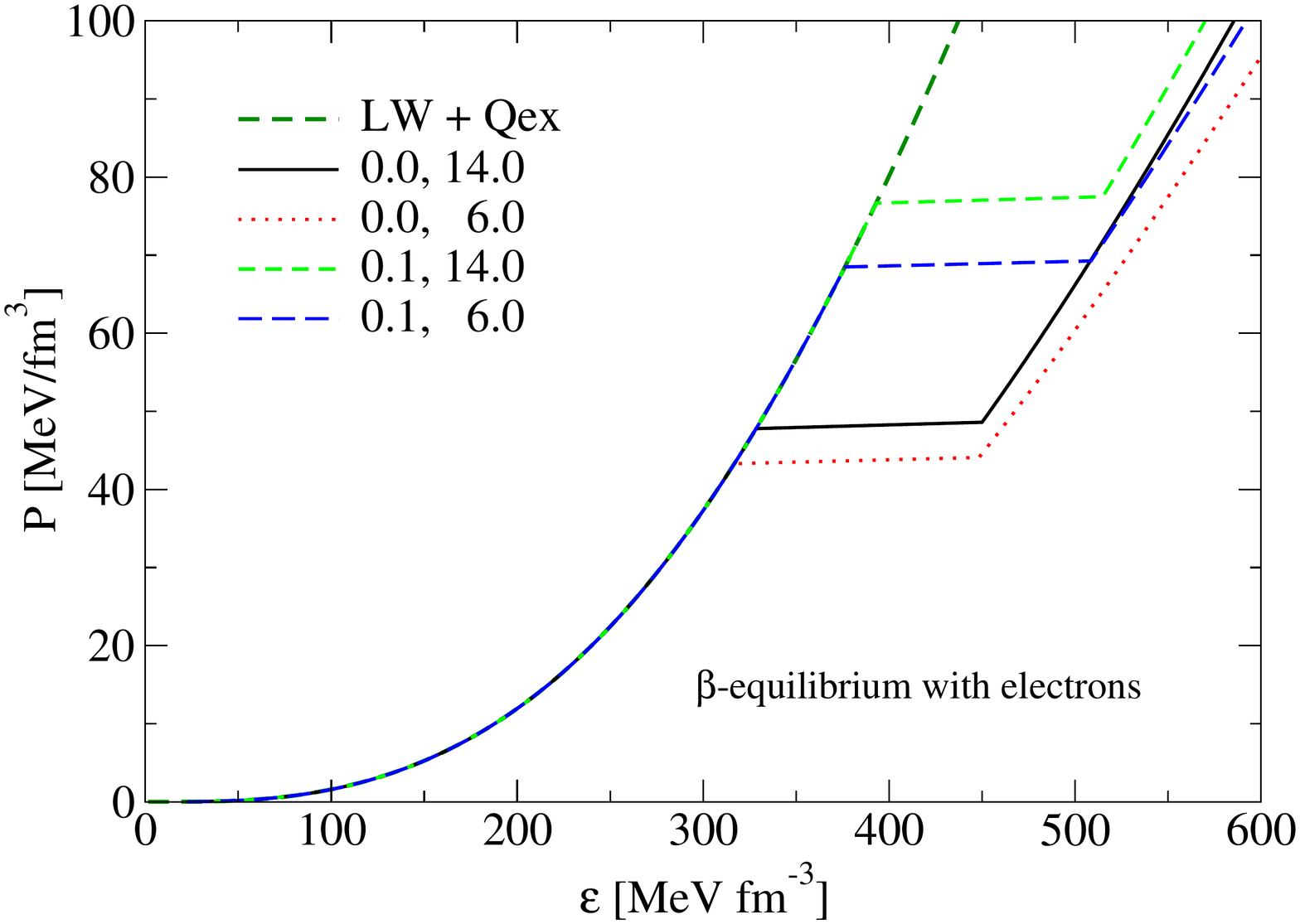}\\{}
	\vspace{-10mm}
	
	\includegraphics[scale=0.3]{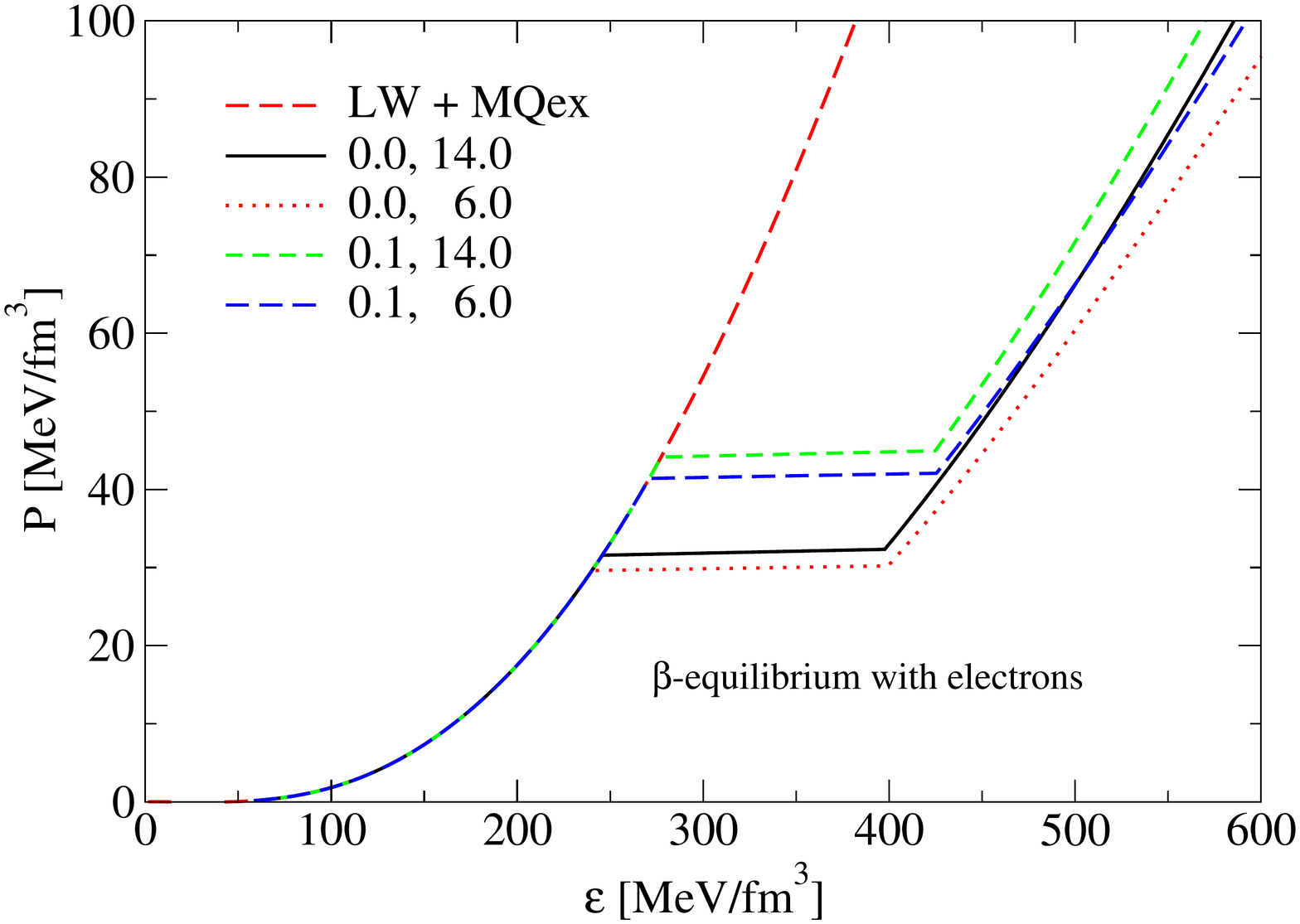} 
	\includegraphics[scale=0.3]{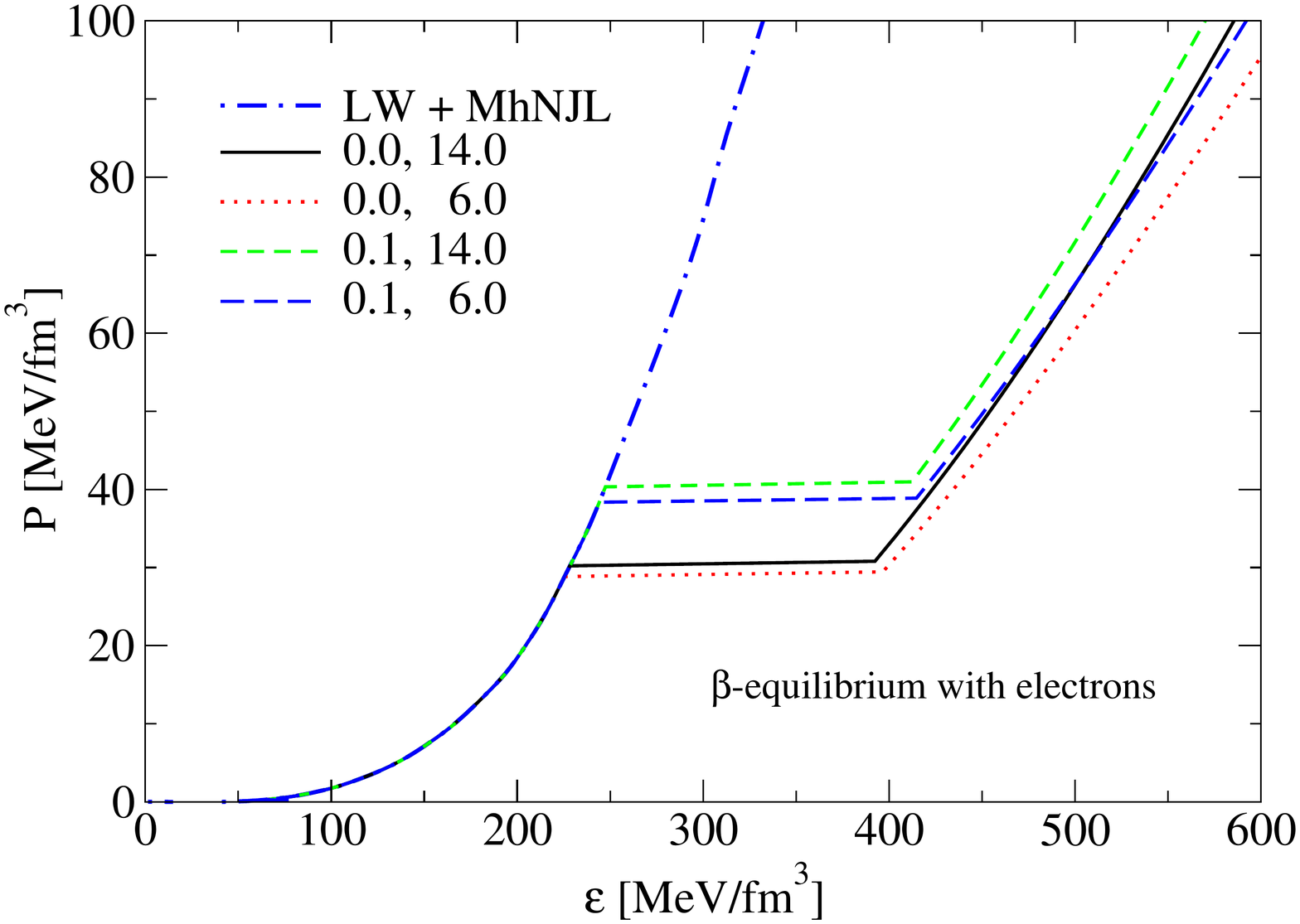}
	\caption{Maxwell construction of the first-order phase transition in the pressure - chemical potential plane.
	The crossing points of the parametrizations of the hNJL quark matter model with the hadronic EoS models define 
	the values for the critical pressure and the critical chemical potential where the system switches from the 
	hadronic to the quark matter phase, described by the corresponding EoS. }
	\label{Maxwell}
\end{figure}

Having defined the hadronic EoS with three different scenarios of the chiral enhancement of the quark Pauli blocking effect,
and four choices for pair of free parameters of the quark matter EoS:  $(\eta_2, \eta_4)= (0, 14.0), (0, 6.0), (0.1,14.0)$  and $(0.1,6.0)$,
we perform four Maxwell constructions for each hadronic model, see Fig.~\ref{Maxwell}.
In the upper left panel of that figure we illustrate the Maxwell construction in the pressure - chemical potential plane.
Each crossing point of a hadronic EoS $P_H(\mu)$ with a quark matter one $P_Q(\mu)$ 
fulfills the Gibbs conditions for phase equilibrium at $T=0$ because the chemical potentials are equal  
to the critical value $\mu_c$ (chemical equilibrium) where the pressures coincide  $P_H(\mu_c)=P_Q(\mu_c)$  
(mechanical equilibrium). 
According to the principles of equilibrium thermodynamics,  the system is at each value of the chemical potential in the 
phase with the highest pressure. Therefore, at the crossing point $\mu_c$ the system switches from the hadronic to the quark
matter EoS. Since at $\mu_c$ the corresponding pressures have a different slope, this transition 
is accompanied with a jump in the baryon number density $n=d P/d\mu$ and energy density $\varepsilon= -P + \mu n$.
 
In the remaining three panels of Fig.~\ref{Maxwell}
we show the pressure as a  function of the baryon density for the 12 hybrid EoS models resulting from the combination
of the three hadronic EoS: LW+Qex (upper right panel), LW+MQex (lower left panel) and LW+MhNJL (lower right panel) 
with the four quark matter EoS for the model parameters of the hNJL model: $\eta_{2}=\{0.0,0.1\}$
and $\eta_{4}=\{6.0,14.0\}.$ 
The other parameters of the hNJL model are fixed to values of Ref.~\cite{Kashiwa:2006rc}, see also section \ref{ssec:hNJL} above.
%$gs_{1}=631.5$, $gs_{2}=2.104$ ,$m_{0}=5.5$MeV.

\begin{figure}[!ht]
\vspace{-10mm}
	\includegraphics[scale=0.45]{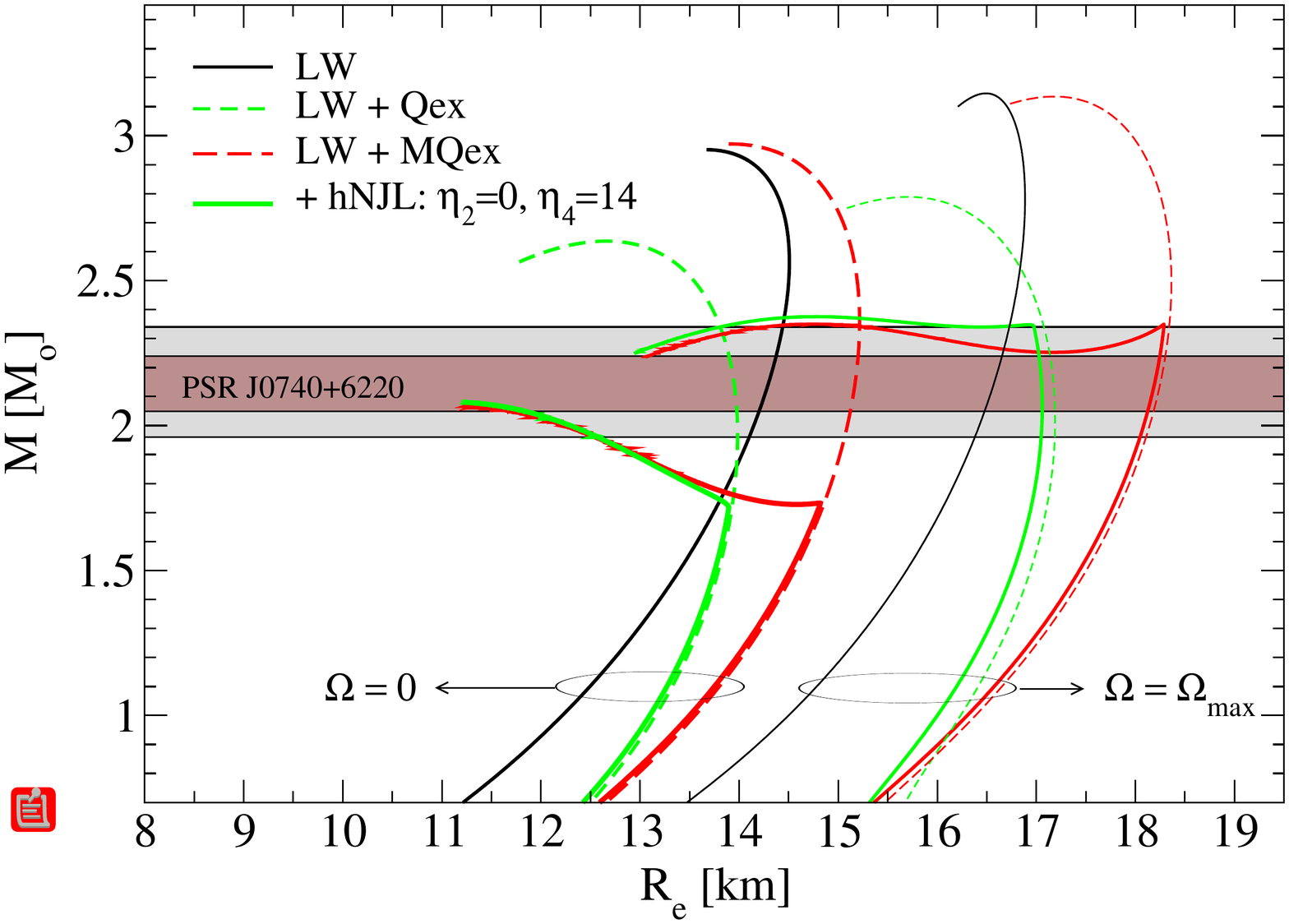}
	\caption{Mass-radius relation for neutron stars within the LW model (black dotted lines) modified by quark exchange effects 
	with two different schemes for the density dependence of the quark mass: constant quark mass (LW+Qex, green short-dashed line), 
	Brown-Rho scaling (LW+MQex, red long-dashed line).	
	The bold lines are for non-rotating star sequences and the thin lines for the rigidly rotating ones with maximal angular velocity.
	Results for a deconfinement phase transition to hNJL quark matter core sequences are shown by solid lines.}
	\label{NSconf}
\end{figure}

In Fig.~\ref{NSconf} we show the mass-radius relation for compact
star configurations considering two models for the density dependence of the quark mass: 
constant mass (LW+Qex) and Brown-Rho scaling (LW+MQex), without and with the possible
phase transition to hNJL quark matter. 
We do not show the M-R curves for LW+MhNJL here because, as we mentioned earlier when
discussing Fig.~\ref{BeqEoS}, this scenario violates causality ($c_s^2 > 1$) at large densities in the hadronic 
phase and makes sense only with a phase transition that prevents this problem to occur.
The phase transition, however, is the same for LW+MQex and LW+MhNJL, 
so that the lines for the latter results are indistinguishable from those for the former ones
and are not displayed separately.
For the hybrid EoS we choose the quark matter model with the parameters 
$\eta_{2}=0$ and $\eta_{4}=14$. 
As it is shown in Fig.~\ref{NSconf}, the differences between models for the masses of stars 
are small, and all of them satisfy the $2~M_{\odot}$ observational constraint from the 
Shapiro-delay based mass measurement on PSR J0740+6220 \cite{Cromartie:2019kug}.

Moreover, with this particular hybrid EoS the third family of compact stars \cite{Gerlach:1968zz}
is possible because the three conditions are fulfilled \cite{Alvarez-Castillo:2016wqj}:
(i) a sufficiently stiff hadronic EoS, (ii) a large jump in energy density at the transition which occurs at a low pressure $P(\mu_c)<100$ MeV/fm$^3$, 
(iii) a sufficiently stiff quark matter EoS to reach a maximum mass of $\sim 2M_\odot$.
Such a third family of compact stars if it would be discovered, would signal a strong first-order phase transition
and therefore support the existence of a critical endpoint in the QCD phase diagram  \cite{Blaschke:2013ana,Alvarez-Castillo:2016wqj}.
Recently, it was shown within a Bayesian analysis that the existence of such a class of hybrid EoS is in accordance 
with modern constraints from multi-messenger astronomy \cite{Blaschke:2020qqj}.  

We like to remark that a similar calculation, with quark Pauli blocking as a repulsive interaction in the nuclear matter 
phase (for constant quark mass) and with the string-flip model for quark matter has been performed as early as
in 1989 with a similar result that stable hybrid stars with quark matter core are possible and have a maximum 
mass above $2~M_\odot$ \cite{Blaschke:1989nn}. 
At that time, measured pulsar masses were below $1.5~M_\odot$. 

In the same plot we show also the relationship between mass and equatorial radius  for
stars rotating with the maximum possible angular velocity. 
These calculations have been performed within the slow-rotation approximation described 
in detail in Refs. \cite{Chubarian:1999yn,Blaschke:2019tbh}

\section{Conclusions}

The relativistic mean field model of the nuclear matter equation of state has
been modified by including the effect of Pauli-blocking owing to quark exchange 
between the baryons. 
Different schemes of a chiral enhancement of the quark Pauli blocking due to a density-dependent
reduction of the value of the dynamical quark mass have been considered.
The resulting equations of state for the pressure have been compared to the RMF model DD2 with 
excluded volume correction.

On this basis a density-dependent nucleon excluded volume is extracted which 
parametrises the quark Pauli blocking effect in the respective scheme of chiral enhancement. 
The dependence on the isospin asymmetry of the quark Pauli blocking is investigated and the corresponding 
density dependent nuclear symmetry energy is obtained in fair accordance with 
phenomenological constraints.

The deconfinement phase transition is obtained by a Maxwell construction 
with a quark matter phase described within a higher order NJL model.
Solutions for rotating and nonrotating (hybrid) compact star sequences are
obtained which show  the effect of high-mass twin compact star solutions for the rotating case. 
This result is a consequence of the stiffening of the nuclear equation of state due to the 
quark Pauli blocking effect which at the same time is a precursor of the delocalisation of the
quark wave function in the deconfinement transition that leads to a strong softening and thus 
a large enough density jump at the phase transition to induce a gravitational instability as one of the  
necessary conditions for the occurrence of a third family solution for hybrid star sequences 
in the neutron star mass-radius diagram. The other one is the sufficient stiffness of 
deconfined quark matter at high densities which is provided by the 8-quark interactions in 
the scalar and vector channels of the higher order NJL model.

\subsection*{Acknowledgments}
The authors thank K. Bugaev for discussions and comments to the manuscript. 
	H.G. acknowledges 
	%support by the DAAD partnership program between the 
	%University of Rostock and the Yerevan State University and 
	the hospitality that was extended to him during his visits at the University of Rostock and at the University of Wroclaw.
	%partner Institution. 
	This work was supported by the Russian Science foundation under grant number 17-12-01427.
%%%%%%%%%%%%%%%%%%%%%%%%%%%%%%%%%%%%%%%%%%%%%%%%%%%%%%
	
\pagebreak

\appendix
\section{Pauli quenching for nucleons in nuclear matter - a quark substructure effect}

At the present there is a growing interest to understand 
the properties of nuclear matter on the basis of the underlying 
quark substructure. 
As long as a first principle QCD-approach to
this problem cannot be realised, semi-phenomenological quark
potential mode approaches can be successfully applied to work out the
description of hadronic properties within a quark picture.
Nonrelativistic quark models have been proven remarkably useful in
describing the hadron spectroscopy \cite{Isgur:1978xj,Isgur:1978wd,Isgur:1979be}. 
Many efforts have been
made to derive the hadron-hadron interaction from say the six-
quark problem. Phase shifts obtained from a non-relativistic quark
potential model give a good fit to the scattering data of the nucleon-nucleon 
\cite{Barnes:1993nu,Oka:1980ax,Oka:1981ri,Oka:1981rj,Burov:1981qh,Faessler:1982ik,Faessler:1983yd,Oka:1985vg}
meson-nucleon  \cite{Barnes:1992ca}
and meson-meson 
\cite{Oka:1984yx,Barnes:1991em,Barnes:1992qa,Blaschke:1992qa} interaction.

Another interesting problem is the investigation of nuclear matter
as a many quark system at finite temperature and density. As a
consequence of their quark substructure the nucleons are affected
by the surrounding nuclear medium. In contrast to the few-quark
problem, where we have to solve the Schr\"odinger equation for 
the isolated three-quark system, a quantum 
statistical approach is needed to treat the many-quark system
at finite temperature. Because of the confinement property of the
quark interaction potential, this quantum statistical approach
must be modified if compared with
usual classical many-particle systems.

To formulate the Hamiltonian we consider non-relativistic massive
quarks so that the kinetic energy is given by
\begin{eqnarray}
\label{KE}
{\rm KE} = \sum_{i=1}^N \left(m+\frac{p_i^2}{2m}\right).
\end{eqnarray}
The potential energy
PE$({\bf r}_1 \dots {\bf r}_N)$
is constructed in the following way \cite{Ropke:1986qs}. 
The configuration $({\bf r}_1 \dots {\bf r}_N)$
is decomposed into color-neutral clusters $q \bar q$ or $qqq$, respectively. 
The confining two-body interaction among quarks is assumed here in the 
form of a harmonic oscillator potential
\begin{equation}
\label{Vij}
 V_{ij}= \frac{m \omega^2}{2}({\bf r}_i - {\bf r}_j)^2
\end{equation}
and shall act only within these color-neutral clusters (saturation property of the interaction). 
Within all possible decompositions 
of the quark configurations one has to take the cluster
configuration with the minimum potential energy, this minimum value
of the potential energy will be denoted by PE.
The Hamiltonian is then given by
\begin{eqnarray}
H = {\rm KE} + {\rm PE} ~.
\end{eqnarray}
Of course, this Hamiltonian is able to describe isolated hadrons
where the quark interaction is confined within the color-neutral
hadronic cluster. 
With respect to the two-nucleon problem \cite{Oka:1985vg},
color van-der-Waals forces do not arise because of the saturation
property of the quark interaction \cite{Horowitz:1985tx}. 
A massive quark matter phase can be described where the potential energy is given by the distribution
function of the next neighbors \cite{Blaschke:1984yj,Schulz:1987qg}.

We consider nuclear matter as the hadronized phase where the
interaction strings are confined within the nucleons and string
flips like in the quark matter phase are not likely to occur.
However, the color-neutral three-quark cluster is influenced by the
surrounding clusters by reason of the Pauli principle what demands
the antisymmetrisation of the hadronic quark wave functions. 
The corresponding shift of the nucleon energy which may be considered
as the self-energy of the three-quark cluster should contribute to
the binding energy of nuclear matter. 
It is the aim of this appendix to evaluate this self-energy contribution due to 
Pauli blocking and to provide it in a simple analytic form that can be 
used in phenomenological approaches in order to account for this 
quark substructure effect when comparing with empirical values 
for the properties of nuclear matter.

Within a Green function approach \cite{Ropke:1986qs} the lowest order diagram
with respect to the density gives the Pauli shift of the three-quark cluster
\begin{eqnarray}
\label{DEnPauli}
 && \Delta E_n^{\rm Pauli}=\sum_{n'} \Delta E_{nn'}^{\rm Pauli}f_3(E_{n'}),
\nonumber \\&&
\Delta E_{nn'}^{\rm Pauli}=3\sum_{1\dots 6} \psi^*_{n}(123) \psi^*_{n'}(456)
\left( KE - E_{n}-E_{n'}\right)
%\nonumber \\&&\times 
\left[\psi_{n}(126) \psi_{n'}(453)-\psi_{n}(453) \psi_{n'}(126)\right].
\end{eqnarray}
This Pauli blocking shift has already been evaluated for finite
temperatures and densities of the nuclear environment and leads to
temperature and density dependent nucleonic properties, such as the
effective nucleon mass \cite{Ropke:1986mh},
%effective nucleonic radii [13] 
and corresponds to the hard-core part of the effective Skyrme interaction 
for nuclear matter \cite{Ropke:1986qs}.

However, at zero temperature the Pauli quenching shift (\ref{DEnPauli}) obtained 
within a quantum statistical treatment of the completely
hadronized quark plasma may be interpreted as a contribution due to
an appropriately chosen antisymmetrisation of the six-quark wave
function $\Phi_{nn'} (1 \dots 6)$ of the two-nucleon problem. 
In this appendix we want to show this correspondence in detail thus coacting
few-body approaches which deal with the problem of effective 
NN-interactions on the quark level using the resonating-group method
\cite{Burov:1981qh,Faessler:1982ik,Faessler:1983yd,Storm:1985pj,Oka:1981ri,Oka:1981rj}.

In the spirit of a perturbation theory, we want to represent the six-quark wave function $\Phi_{nn'} (1 \dots 6)$  as a
product of two nucleonic wave functions that behaves antisymmetrically
with respect to each exchange of quantum numbers belonging to quarks
($P_{i j}; i=1,2,3; j=4,5,6$) or to nucleons ($P_{nn'}$) thus fulfilling the
Pauli principle on the nucleonic as well as on the quark level. 
Following this prescription and considering only the two~nucleon channel, all those permutations leading to
color non-singlet clusters have to be excluded and we obtain
\begin{equation}
\label{Asymm}
 \Phi_{nn'} (1 \dots 6)=\left(1-\sum_{i=1}^3 P_{i,i+3} \right) \left(1- P_{nn'} \right) \psi_{n}(123) \psi_{n'}(456)\,,
\end{equation}
where the numbers $i=1  \dots 6$ stand for the momentum, spin, flavor
and color indices of the $i$-th quark and $n$ denotes the center-of-mass momentum $\bf P$
as well as one of the spin-isospin orientations
of the ground state nucleon ($\nu =p \uparrow, p \downarrow,n \uparrow , n \downarrow$). The wave function
$\psi_{n} (123)$ of the nucleon can be found as the ground state solution
of the three-quark Hamiltonian
\begin{equation}
\label{Hamil}
 H(123)=\sum_{i=1}^3 (m+\frac{p_i^2}{2 m})+\sum_{i < j=2}^3 V_{ij}
\end{equation}
with the harmonic oscillator confinement potential (\ref{Vij}). Since
the Hamiltonian (\ref{Hamil}) is independent of spin, flavor and color (SFC)
of the constituent quarks, the SFC-part $\chi_\nu (123)$ can be separated
from the orbital part $\varphi_P(123)$ of the nucleon wave function
according to 
\begin{equation}
\label{WF}
\psi_{n} (123)=\varphi_P(123) \chi_\nu (123) .
\end{equation} 
The property of antisymmetry of the three-quark wave function
determines the symmetry properties of the $\varphi_P$
and the $\chi_\nu $ part.
In a systematic way, this decomposition can be done by using the
technique of Young tableaux. The lowest energy eigenvalue corresponds 
to a total symmetric orbital part, with respect to spin and
flavor the wave function has a mixed symmetry, whereas for the
color part a total antisymmetric function is needed, see also [10].
With explicit notation of the spin ($\uparrow,
\downarrow$), flavor ($u, d$) and
color ($R,G,B$) degrees of freedom, the SFC-part of the nucleon
wave function reads
\begin{eqnarray}
\label{SFC}
 \chi_\nu (123)&=&\frac{1}{\sqrt{18}}\left(2 u \uparrow  u \uparrow d \downarrow
+ 2 u \uparrow d \downarrow u \uparrow  +2 d \downarrow u \uparrow  u \uparrow 
\right. \nonumber \\ && \left.
-u \uparrow u \downarrow d \uparrow -u \uparrow d \uparrow u \downarrow  
-d \uparrow u \uparrow u \downarrow -u \downarrow u \uparrow  d \uparrow 
\right. \nonumber \\ && \left.
-u \downarrow d \uparrow u \uparrow - d \uparrow u \downarrow u \uparrow \right) 
\frac{1}{\sqrt{6}} \det |RGB|.
\end{eqnarray}

By alternating the spin or isospin orientations in (\ref{SFC}), the
four species of ground state nucleons ($\nu =p \uparrow, p \downarrow,n \uparrow , n \downarrow$) 
are described. 
The orbital part of the nucleonic wave function is obtained by solving the Schr{\"o}dinger equation
\begin{equation}
 H(123) \varphi_P(123) =E_n \,\varphi_P(123) ; \qquad n=P, \nu,
\end{equation}
yielding for the ground state
\begin{equation}
\label{phi_P}
 \varphi_P(123) =\frac{8 \pi^3}{V}\left(\frac{\sqrt{3} b^2}{\pi}\right)^{3/2}
\delta_{{\bf P},{\bf P}_R} e^{-(p^2_\rho+p^2_\lambda)b^2/2},
\end{equation}
\begin{equation}
\label{En}
 E_n =P^2/6m+3m+3 \sqrt{3} \omega.
\end{equation}
Here we have used the Jacobi coordinates
\begin{eqnarray}
\label{Jacobi}
{\bf P}_R&=&{\bf p}_1 +{\bf p}_2+{\bf p}_3,\nonumber \\
{\bf p}_\rho&=&\frac{1}{\sqrt{2}}({\bf p}_1 -{\bf p}_2),\nonumber \\
{\bf p}_\lambda&=&\frac{1}{\sqrt{6}}({\bf p}_1 +{\bf p}_2-2{\bf p}_3),
\end{eqnarray}
and the width parameter of the gaussian wave function
$b^{-2}=\sqrt{3} m \omega; \,\,\hbar = 1$, $V$ is the normalization volume.

Now, the antisymmetrized two-nucleon wave function follows from
(\ref{Asymm}) with (\ref{WF}),(\ref{SFC}) and (\ref{phi_P}). 
The normalization is given by
\begin{eqnarray}
 N_{nn'}&=& \langle \Phi_{nn'}|\Phi_{nn'} \rangle
 \nonumber \\
 &=&1-\delta_{{\bf P},{\bf P}'}
-3 \sum_{p_1 \dots p_6} \varphi^*_P(123)\varphi^*_{P'}(456)
%\nonumber \\ &&\times 
\left[ c^{(1)}_{\nu\nu'}\varphi_P(126)\varphi_{P'}(453)+
c^{(2)}_{\nu\nu'}\varphi_P(453)\varphi_{P'}(126)\right].
\label{Nnn}
\end{eqnarray}
Here, the $c^{(1)}_{\nu\nu'}$ and $c^{(2)}_{\nu\nu'}$  reflect the scalar products 
of the SFC-part with exchange according to $P_{3,6}$ and $P_{nn'}P_{3,6}$
\begin{eqnarray}
 c^{(1)}_{\nu\nu'}&=& \langle \chi_\nu (123)\chi_{\nu'} (456)
\chi_\nu (126)\chi_{\nu'} (453) \rangle \nonumber\\
 c^{(2)}_{\nu\nu'}&=& - \langle \chi_\nu (123)\chi_{\nu'} (456)
\chi_\nu (453)\chi_{\nu'} (126) \rangle .
\end{eqnarray}
The color degrees of freedom are immediately elaborated by
rearranging the color variables, a factor 2 arises from two
different variants of $\chi_\nu$ if the non exchanged variables are
transposed. The remaining SF-variables are explicitly written
down and evaluated. 
The results for $\nu= n \uparrow$ are given in table \ref{tab:1},
the equivalent results hold also for the other nucleon states,
if the interaction is invariant with respect to the isospin
variables.

The momentum variables are integrated taking into account that
the exchange operator $P_{3,6}$ is different from zero only for
${\bf p}_3={\bf p}_6$. The result can be given in a closed form
\begin{eqnarray}
  N_{nn'}&=& 1-\delta_{{\bf P},{\bf P}'}-\frac{9 \sqrt{3}}{8}
\left(\frac{b^2}{\pi}\right)^{3/2} \frac{8 \pi^3}{V}
%\nonumber \\&& \times 
\left[ c^{(1)}_{\nu\nu'} e^{-({\bf P}-{\bf P}')^2 b^2/12}
+c^{(2)}_{\nu\nu'} e^{-({\bf P}-{\bf P}')^2 b^2/3}\right].
\end{eqnarray}

%Table 1. The values of the matrix elements $c_{\nu\nu'}$ and $d_{\nu\nu'}$.

\begin{table}[htbp]
\caption{The values of the matrix elements $c^{(1)}_{\nu\nu'}$ and $c^{(2)}_{\nu\nu'}$
for $\nu = n \uparrow$. }
\begin{center}
\begin{tabular}{|c|c|c|c|}
\hline
$\nu$ &$\nu'$ &$c^{(1)}_{\nu\nu'}$ &$c^{(2)}_{\nu\nu'}$ \\
\hline
$n \uparrow$ & $n \uparrow$ & 31/243 & - 31/243 \\
$n \uparrow$ & $n \downarrow$ & 14/243 & - 17/243 \\
$n \uparrow$ & $p \uparrow$ & 14/243 & - 17/243 \\
$n \uparrow$ & $p \downarrow$ & 22/243 & - 25/243 \\
\hline
&$\sum_{\nu'}$ & 1/3 & - 10/27 \\
\hline
\end{tabular}
\end{center}
\label{tab:1}
\end{table}%

%Evaluation of the Energy Shift

The antisymmetrisation of the two-nucleon wave function with
respect to the quark degrees of freedom leads to a shift in the
two-nucleon energy according to
\begin{equation}
\label{DEnnPauli}
 \Delta E_{nn'}^{\rm Pauli}=
\frac{1}{ N_{nn'}} \langle \Phi_{nn'}|H|\Phi_{nn'} \rangle -E_n-E_{n'},
\end{equation}
with $E_n$ given by Eq. (\ref{En}). 
The Hamiltonian $H= KE+PE$ contains the kinetic part, Eq. (\ref{KE}), 
and the potential part, Eq. (\ref{Vij}).
Neglecting the antisymmetrisation of the wave function with respect
to quark exchange, the kinetic energy which is in Jacobi coordinates
\begin{equation}
 KE= 6m+\frac{P^2_R}{6m}+\frac{{P'}^2_R}{6m}+\frac{1}{2m} 
 \left(p^2_\rho +p^2_\lambda+{p'}^2_\rho+{p'}^2_\lambda \right),
\end{equation}
and the potential energy
\begin{equation}
 PE=3 \frac{m \omega^ 2}{2} \left(\rho^2+\lambda^2+{\rho'}^2+{\lambda'}^2\right),
\end{equation}
are immediately evaluated for the two-nucleon system with the result
\begin{equation}
  \langle \Phi_{nn'}|KE|\Phi_{nn'} \rangle\approx  6m+\frac{P^2_R}{6m}
  +\frac{{P'}^2_R}{6m}+3 \sqrt{3} \omega,
\end{equation}
\begin{equation}
 \langle \Phi_{nn'}|PE|\Phi_{nn'} \rangle\approx   3 \sqrt{3} \omega,
\end{equation}
so that no energy shift arises.

Orthogonalisation of the wave function by antisymmetrisation will
lead to a change in the kinetic energy. In contrast to the kinetic
energy, the potential energy (\ref{Vij}) is not determined by the wave
function but by the density distribution of the quarks. 
In particular, the probability of a given quark configuration is determined 
by the density distribution function. It is well-known from
the Hartree-Fock theory that antisymmetrisation will not change
the particle density distribution 
$\rho({\bf r})=\sum_i \delta({\bf r}-{\bf r}_i)$. 
For two nucleons we obtain an overlap of the quark density distributions,
and the potential energy in not significantly changed by the antisymmetrisation procedure, 
as long as string flip processes are not of importance. As discussed below, a variation of the wave
function beyond the scope of a Hartree-Fock type antisymmetrisation will also lead to a variation of the potential energy.

In this way, the energy shift (\ref{DEnnPauli}) is determined by the change of the kinetic energy with
\begin{equation}
\label{DEnn'Pauli}
\Delta E_{nn'}^{\rm Pauli}=N_{nn'}^{-1}\left(E_n+E_{n'}+ \Delta KE_{nn'}^{\rm Pauli}\right)
 -E_n-E_{n'},
\end{equation}
with
\begin{eqnarray}
 \Delta KE_{nn'}^{\rm Pauli}&=& -(E_n+E_{n'})
-3 \sum_{p_1 \dots p_6} \varphi^*_P(123)\varphi^*_{P'}(456)
\nonumber \\ &&\times 
KE \left[ c^{(1)}_{\nu\nu'}\varphi_P(126)\varphi_{P'}(453)+
c^{(2)}_{\nu\nu'}\varphi_P(453)\varphi_{P'}(126)\right].
%\nonumber \\
\end{eqnarray}
Expanding the normalisation factor $N_{nn'}^{-1}$ up to the first order
with respect to the overlap integral, see Eq. (\ref{Nnn}), the expression
(\ref{DEnPauli}) for the Pauli shift is recovered.

The interpretation of the energy shift due to the Pauli blocking
can be given in correspondence to atomic physics. 
At short interatomic distances, the energy of the two-atom system is sharply
increasing what is usually represented by a repulsive, hard core
like interaction potential. 
Indeed, the physical reason of this increase of energy is not the 
Coulombic electron-electron interaction, but the increase of kinetic energy 
because of the Pauli principle which demands the orthogonalisation of the electron
wave functions.

Now, let us proceed to the explicit evaluation of the Pauli-blocking 
shift (\ref{DEnn'Pauli}) which may be given using the Jacobi coordinates (\ref{Jacobi}) as follows:
\begin{eqnarray}
&& \Delta E_{nn'}^{\rm Pauli}=3 \sum_{P_R,p_\rho,p_\lambda} 
 \sum_{P'_R,p'_\rho,p'_\lambda} \frac{\partial (p_1 \dots p_6)}{\partial(P_R\dots p'_\lambda)}
\delta_{{\bf P},{\bf P}_R} \delta_{{\bf P}',{\bf P}'_R}
%\nonumber \\
\left[6 \sqrt{3} \omega -\frac{1}{2m} \left(p^2_\rho+p^2_\lambda+{p'}^2_\rho
+{p'}^2_\lambda \right) \right]\nonumber \\
&&\quad\times \left[c^{(1)}_{\nu\nu'}\delta_{{\bf P},{\bf P}_R-({\bf P}_R-{\bf P}'_R)/3+2 ({\bf p}_\lambda -{\bf p}'_\lambda)/\sqrt{6}}
+ c^{(2)}_{\nu\nu'}\delta_{{\bf P},{\bf P}_R+({\bf P}_R-{\bf P}'_R)/3 -2 ({\bf p}_\lambda-{\bf p}'_\lambda)/\sqrt{6}}\right]
%\nonumber \\
{\rm e}^{ -b^2 \left(p^2_\rho+p^2_\lambda+{p'}^2_\rho +{p'}^2_\lambda \right)}
\nonumber \\
&& = \frac{9 \sqrt{3}}{16} \frac{8 \pi^3}{V}
\left(\frac{b^2}{\pi}\right)^{3/2} \frac{1}{mb^2}
\left\{ c^{(1)}_{\nu\nu'} e^{-({\bf P}-{\bf P}')^2 b^2/12} \left[\frac{15}{2}-\frac{b^2}{12} ({\bf P}-{\bf P}')^2 \right]
\right. \nonumber\\&&\left. \hspace{4.5cm}
+c^{(2)}_{\nu\nu'} e^{-({\bf P}-{\bf P}')^2 b^2/3} \left[\frac{15}{2}-\frac{b^2}{3} ({\bf P}-{\bf P}')^2 \right] \right\}.
\label{DEnn'Paul}
\end{eqnarray}

\begin{figure}[!h]
	\includegraphics[scale=0.35]{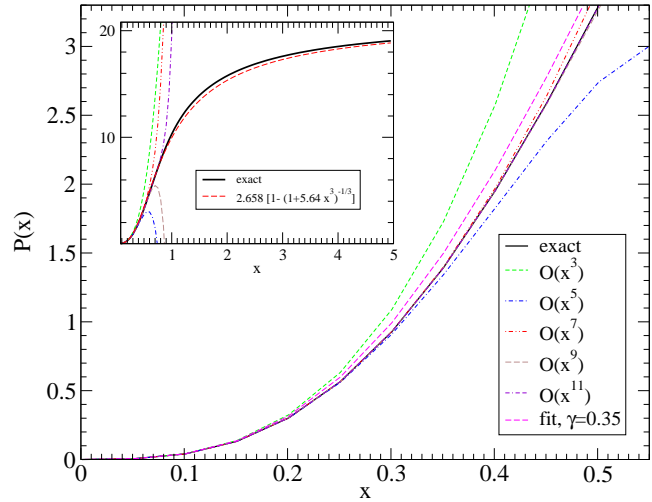}
	\caption{
%	Contributions of one ($\alpha=1$, solid line) and two ($\alpha=2$, dashed line) quark exchange terms in self energy.
The function P(x) defined in Eq.~(\ref{eq:P-x})	with its polynomial expansions in different lowest orders, as well as a fit that obeys 
both limits of $P(x\to 0) = 40 x^3$ and $P(x\to \infty) = 12\sqrt{\pi}$. In the main panel the range of applicability of the low-density approximation 
is shown, a larger picture is shown in the inset.
} 
	\label{walpha} 
\end{figure}

Whereas this quantity (\ref{DEnn'Paul}) measures the surplus energy arising
from the antisymmetrisation of the wave function with respect
to the two-nucleon problem, we are especially interested in the
energy shift for a single nucleon $\Delta E_n^{\rm Pauli}$ in a many-nucleon
system which can be obtained from (\ref{DEnn'Paul}) by summation over the
second nucleonic index $n'$, whereby at $T=0$ the respective distribution function 
(see (\ref{DEnPauli})) is a step function restricting the
momentum summation to the range of the Fermi sphere $|{\bf P}'| < P_F$.
The sum over ${\bf P}'$ may then be evaluated as an integral yielding
\begin{eqnarray}
\Delta E_{\nu P}^{\rm Pauli}(P_{F,n},P_{F,p})&=& 
\sum_{\nu'}V\int_{|{\bf P}'| < P_F} \frac{d^3 {\bf P}'}{(2 \pi)^3} \Delta E_{nn'}^{\rm Pauli}
= \sum_{\tau'=n,p}\sum_{\alpha=1,2}c_{\tau \tau'}W_\alpha (P,P_{F,\tau'})\\
%\nonumber \\
W_\alpha(P,P_{F})&=&\overline{W}_\alpha \lambda_\alpha^3 \int_0^{P_F} dP' \,P'^2 \int_{-1}^1 dz \left\{
e^{-\lambda_\alpha^2(P^2+{P'}^2-2P\,P'z)} \left[\frac{15}{2}-\lambda_\alpha^2 (P^2+{P'}^2-2P\,P'z)\right] \right\} ~,
\nonumber \\ 
\end{eqnarray}
where the abbreviations
$\overline{W}_\alpha=\frac{9 \sqrt{3}}{64 \sqrt{\pi}} \frac{b}{m}/\lambda_\alpha^3$ and $\lambda_\alpha=\frac{\alpha}{2\sqrt{3}}b$
have been used. We introduce dimensionless momenta $x_\alpha=\lambda_\alpha P$ and analogous for the primed momentum
as well as the Fermi momentum and perform the angular integration over the $z$-variable

\begin{eqnarray}
W_\alpha(x_\alpha,x_{\alpha,F})&=&\overline{W}_\alpha \frac{1}{x_\alpha} \int_{0}^{x_{\alpha,F}} dx_\alpha^\prime\,x_\alpha^\prime 
\left\{{\rm e}^{-(x_\alpha-x_\alpha^\prime)^2} \left[\frac{13}{2}-(x_\alpha-x_\alpha^\prime)^2\right] 
-{\rm e}^{-(x_\alpha+x_\alpha^\prime)^2} \left[\frac{13}{2}-(x_\alpha+x_\alpha^\prime)^2\right] \right\} 
\nonumber \\ 
&=& \overline{W}_\alpha \frac{1}{x_\alpha} \int_{-x_{\alpha,F}}^{x_{\alpha,F}} dx_\alpha^\prime\,x_\alpha^\prime
\left\{{\rm e}^{-(x_\alpha-x_\alpha^\prime)^2} \left[\frac{13}{2}-(x_\alpha-x_\alpha^\prime)^2\right] \right\} 
\nonumber \\ 
&=& \overline{W}_\alpha \frac{1}{x_\alpha} \int_{-x_{\alpha}-x_{\alpha,F}}^{-x_{\alpha}+x_{\alpha,F}} dx_\alpha^\prime\,
{\rm e}^{-{x_\alpha^\prime}^2}\left(\frac{13}{2}-{x_\alpha^\prime}^2\right)(x_\alpha^\prime+x_\alpha)
\nonumber \\ 
&=& \overline{W}_\alpha\left\{12 \sqrt{\pi} \left[ {\rm erf} (x_{\alpha,F}-x_{\alpha}) -  {\rm erf} (x_{\alpha,F}+x_{\alpha})\right]\right.
\nonumber \\ 
&&\left. + \frac{{\rm e}^{-(x_{\alpha,F}+x_\alpha)^2}}{x_{\alpha}}  \left[11 - 2 x_{\alpha,F}(x_{\alpha,F}+x_{\alpha})\right]
-  \frac{{\rm e}^{-(x_{\alpha,F}-x_\alpha)^2}}{x_{\alpha}}  \left[11 - 2 x_{\alpha,F}(x_{\alpha,F}-x_{\alpha})\right]\right\}~.
\label{eq:walpha}
\end{eqnarray}
For the applications to symmetric nuclear matter and to pure neutron matter we just need the case that  both arguments are equal 
to the same Fermi momentum, i.e. 
\begin{equation}
\label{eq:walpha2}
W_\alpha(x_{\alpha,F})=W_\alpha(x_{\alpha,F},x_{\alpha,F})=\overline{W}_\alpha P(x_{\alpha,F})~,
\end{equation}
where 
\begin{equation}
\label{eq:P-x}
P(x)=12 \sqrt{\pi}\, {\rm erf} (2x)+ \frac{1}{x} \left[{\rm e}^{-4x^2}(11 - 4 x^2) - 11\right]~,
\end{equation}
is the Pauli blocking function that described the momentum dependence of the quark exchange between three-quark clusters
and is depicted in figure \ref{walpha} together with its power law expansion up to a given order,
\begin{equation}
\label{eq:Ppower}
P(x)=40 x^3 - \frac{1088}{15} x^5 + \frac{608}{7}x^7 - \frac{3584}{45} x^9 + \frac{4436}{99} x^{11}- \mathcal{O}(x^{13})~.
\end{equation}
This divergent series is useful only in the low-density (i.e. low-momentum) limit, but it does not display the fact that this function 
asymptotically approaches the constant $12\sqrt{\pi} \approx 21.269$.
%
%%%%%%%%%%%%%%%%%%%%%%%%%%%%%%%%% Figure 1 %%%%%%%%%%%%%%%%%%%%%%%%%%%%%%%
%
%
%%%%%%%%%%%%%%%%%%%%%%%%%%%%%%%%%%%%%%%%%%%%%%%%%%%%%%%%%%%%%%%%%%%%%
An excellent fit to the exact result (\ref{eq:P-x}) is given by 
\begin{equation}
\label{eq:Pfit}
P(x)\approx 12 \sqrt{\pi} \left[1-\left(1+\frac{10}{3\sqrt{\pi}\gamma} x^3\right)^{-\gamma} \right]~~,~\gamma=0.35,
\end{equation}
which is a sufficiently simple function of the nucleon density $n\propto x^3$.
Note that we need to use this function with different arguments for the one-quark ($\alpha=1$)
and two-quark ($\alpha=2$) exchange contributions to the nucleon self energy which have a different range in momentum space.

This function $P(x)$ is used in the main text when the effect of quark Pauli blocking between nucleons on the nuclear equation of state is 
numerically evaluated and discussed.

%\bibliography{references}

\end{document}